\documentclass[showpacs,twocolumn,superscriptaddress]{revtex4}
\UseRawInputEncoding
\usepackage{doi}
\usepackage{hyperref}
\hypersetup{
  colorlinks=true,        
  linkcolor=blue,         
  citecolor=cyan,         
}

\usepackage[utf8x]{inputenc}
\DeclareUnicodeCharacter{2212}{\textendash}
\usepackage{graphicx}
\usepackage{dcolumn}
\usepackage{bm}
\usepackage{color}

\usepackage{amsmath}
\usepackage{amssymb}
\usepackage{mathrsfs}

\begin{document}

\title{Dynamics of particles and epicyclic motions around Schwarzschild-de-Sitter black hole in perfect fluid dark matter}

\author{Javlon Rayimbaev}
\email{javlon@astrin.uz}
\affiliation{Ulugh Beg Astronomical Institute, Astronomy St. 33, Tashkent 100052, Uzbekistan}
\affiliation{Akfa University, Kichik Halqa Yuli Street 17,  Tashkent 100095, Uzbekistan}
\affiliation{National University of Uzbekistan, Tashkent 100174, Uzbekistan}
\affiliation{Institute of Nuclear Physics, Ulugbek 1, Tashkent 100214, Uzbekistan}

\author{Sanjar Shaymatov}
\email{sanjar@astrin.uz}
\affiliation{Ulugh Beg Astronomical Institute, Astronomy St. 33, Tashkent 100052, Uzbekistan}
\affiliation{Akfa University, Kichik Halqa Yuli Street 17,  Tashkent 100095, Uzbekistan}
\affiliation{Institute for Theoretical Physics and Cosmology, Zhejiang University of Technology, Hangzhou 310023, China}
\affiliation{National University of Uzbekistan, Tashkent 100174, Uzbekistan}
\affiliation{Tashkent Institute of Irrigation and Agricultural Mechanization Engineers, Kori Niyoziy 39, Tashkent 100000, Uzbekistan }

\author{Mubasher Jamil} \email{mjamil@zjut.edu.cn (corresponding author)}\affiliation{Institute for Theoretical Physics and Cosmology, Zhejiang University of Technology, Hangzhou 310023, China} \affiliation{School of Natural Sciences, National University of Sciences and Technology, Islamabad 44000, Pakistan}

\date{\today}
\begin{abstract}
In this paper we investigate circular orbits for test particles around Schwarzschild-de Sitter (dS) black hole surrounded by perfect fluid dark matter. We determine the region of circular orbits bounded by innermost and outermost stable circular orbits. We show that the impact of the perfect fluid dark matter shrinks the region where circular orbits can exist as the values of both innermost and outermost stable circular orbits decrease. We find that for specific lower and upper values of dark matter parameter there exist double matching values for inner and outermost stable circular orbits. It turns out that the gravitational attraction due to the dark matter contribution dominates over cosmological repulsion. This gives rise to a remarkable result in the Schwarzschild-dS black hole surrounded by dark matter field in contrast to the Schwarzschild-dS metric. Finally, we study epicyclic motion and its frequencies with their applications to twin peak quasi-periodic oscillations (QPO) for various models. We find corresponding values of the black hole parameters which could best fit and explain the observed twin peak QPO object GRS 1915+109 from microquasars.

\end{abstract}
\maketitle

\section{Introduction}
\label{introduction}

In general relativity black holes have been always very fascinating and intriguing objects for their geometric properties and their existence has been regarded as a generic result of Einstein gravity.  However, they had been so far considered as candidates till the discovery of gravitational waves has been announced as a result of two stellar black hole mergers \cite{Abbott16a,Abbott16b} via LIGO-VIRGO detection and as well the first supermassive M87 black hole image under Collaboration of the Event Horizon Telescope (EHT)~\cite{Akiyama19L1,Akiyama19L6}. After these discoveries the new door has been opened to reveal unknown properties of black hole candidates and to precise constraints and measurements of the parameter related to the geometry of astrophysical black hole candidates.  

Since particles and even photons get affected drastically due to the strong gravitational field of black hole and exotic compact objects their motion has been thoroughly investigated for several years with great interest. In this respect new observations associated with black holes gives new prosperity in understanding not the unexplored problems of black hols candidate but also the other existing fields affecting geodesics of particles and the background geometry. These exiting fields surrounding the black hole candidate may contribute to the motion of particles~\cite{Wald74,Benavides-Gallego19,Herrera00,Herrera05,Bini12,Toshmatov19d}. Thus, the test particle motion can accordingly provide information about the other elements in the black hole vicinity. For example, the magnetic field due to the Lorentz force can drastically affect on the charged particles near the black holes irrespective of the fact that it is  weak~\cite{Prasanna80,Kovar08,Kovar10,Shaymatov14,Shaymatov15,
Dadhich18,Pavlovic19,Shaymatov18a,Shaymatov19b,Stuchlik20,
Shaymatov20egb,Rayimbaev2020PRD,Narzilloev2020EPJC,
Narzilloev20a,BokhariPhysRevD2020,Abdujabbarov2020Galax,
Shaymatov21b,Duztas-Jamil20,Shaymatov21c}. Similarly, in a realistic astrophysical scenario one can also consider the effect arising from the dark matter fields in the background environment of black holes due to the fact that supermassive black holes may be surrounded by dark matter distribution. Although there still exists no direct experimental detection of dark matter, the observational data has verified its existence in galactic rotation curves of giant elliptical and spiral galaxies~\cite{Rubin80}. Note that recent analysis and observations confirm that the contribution of the dark matter to the mass of galaxy can reach up to 90\%~\cite{Persic96}, and hence it is believed that giant elliptical and spiral galaxies endowed with supermassive black holes in their centre are placed in a giant dark matter halos~\cite{Akiyama19L1,Akiyama19L6}. The question then arises--how the background geometry can include dark matter contribution? The question was then addressed by using non-vacuum solution of Einstein's equation by several authors in literature (see for example \cite{Kiselev03,Li-Yang12}). In order to model dark matter background a static and spherically symmetric black hole solution containing a dark matter profile was proposed by Kiselev~\cite{Kiselev03}, containing a logarithmic term, $\alpha\ln(r/r_q)$, as a non-vanishing contribution of dark matter. Later, from a different perspective Li and Yang~\cite{Li12} approached another black hole solution which contains a term $(r_q/r)\ln(r/r_q)$ by imposing the condition, according to which the dark mater halo represented by a phantom scalar field is composed of the weakly interacting massive particles (WIMPs) with the equation of state $\omega\simeq 0$. In doing so, Li and Yang~\cite{Li12} derived the black hole solution which represents only the background matter as a dark matter. Following Li and Yang there are several investigations that can contain dark matter fields in a different way in the black hole background geometry~\cite{Xu18,Schee2019EPJC,Stuchlik2019,Haroon19,Konoplya19plb,
Hendi20,Jusufi19,Shaymatov:2020wtj,Narzilloev2020PhRvD,
Nampalliwar2021arXiv,Rizwan2019PhRvD,Jusufi2020EPJC,Shaymatov20b}.

Similarly to the dark matter field, what to talk of realistic astrophysical scenario, cosmological effects should be taken into consideration at large scales and in black hole vicinity as well. So far we have confirmed that the universe is already being expanded with an accelerating rate due to the unknown repulsive effect of dark energy related to the vacuum energy associated with the cosmological constant term $\Lambda g_{\alpha\beta}$ in the standard Einstein equations. That is, the background of black holes can be no longer considered as vacuum as that of cosmological constant. Thus, the repulsive effect of cosmological constant becomes increasingly important at large scales even though its estimated value is about $\Lambda\sim 10^{-52} m^{-2}$ as stated by recent cosmological observations~\cite{Peebles03,Spergel07}. 
There are many works revealing the significant role of the cosmological constant in wide range of astrophysical phenomena \cite{Stuchlik83,Stuchlik05,Cruz05,Stuchlik11,Grenon10,
Rezzolla03a,Arraut13,Arraut15,Faraoni15,Faraoni16,Shaymatov18a}. Later, the quintessential scalar fields were also suggested instead of the cosmological constant $\Lambda$ as an alternative form of the dark energy \cite{Peebles03,Wetterich88,Caldwell09}. Black hole solutions surrounded by a dynamical quintessential
field were suggested by providing the equation of state $p = \omega_{q}\rho$
with $\omega_q\in(-1;-1/3)$ \cite{Kiselev03} and $\omega_{q}\in(-1;-2/3)$ \cite{Hellerman2001JHEP}. Note that the equation of state with $\omega_{q} = −1$ refers to the vacuum energy with cosmological constant $\Lambda$.

 Generally in the low mass x-ray binary systems such as a neutron star (NS) or a black hole, quasiperiodic oscillations (QPOs) are observed in their power spectra. They are generally characterized by either low frequency (LF) or high-frequency (HF) QPOs. The later ones which usually arise in pairs are therefore termed twin peak HF QPOs. The HF QPOs carry unique information on the matter falling and/or moving in extreme gravity around the compact object. The LF QPOs are strong, persistent, and tend to drift in frequency while HF QPOs are transient and weak but do not shift their frequencies significantly \cite{Tasheva:2018taa,Stefanov:2017bfu}, and in some x-ray binaries, both LF and HF QPOs are found together. The astrophysical data suggests that HF and LF QPOs are created in different parts of the accretion disk. In literature, there are several models available to explain QPOs \cite{Germana:2018snv,Tarnopolski:2021ula,Dokuchaev:2015ghx,Kolos:2015iva,Aliev:2012rj,Stuchlik:2007xt,Titarchuk:2005rr}.

The principal aim of this present paper is to consider these mentioned realistic scenarios and their combined effect to the study of epicyclic motion and its applications to the quasi periodic oscillations (QPOs) around Schwarzschild-de-Sitter black hole surrounded by perfect fluid dark matter. This is what we wish to demonstrate in this paper by studying particle motion in the black hole metric described by the line element proposed in \cite{Li-Yang12,Xu16-dm}.

The paper is organized as follows: In Sec.~\ref{Sec:metic} we briefly introduce the black hole metric and its properties. In Sec.~\ref{Sec:motion} we provide a detailed analysis related to circular orbits of test particles, the degeneracy relation, stability (instability) of circular orbits, and the efficiency the energy released around black hole. In Sec ~\ref{Sec:frequency} we consider fundamental frequencies in the black hole vicinity. We end up our concluding remarks in Sec.~\ref{Sec:conclusion}.

Throughout we use a system of units in which gravitational constant and velocity of light are set to unity. Greek indices are taken to run from 0 to 3, while Latin indices from 1 to 3.

\section{\label{Sec:metic}
Schwarzschild-dS Black hole immersed in perfect fluid dark matter field}

The Lagrange density describing Schwarzschild-de-Sitter black hole immersed in perfect fluid dark matter field is given by \cite{Li-Yang12,Xu16-dm} 
\begin{eqnarray}
{S}&=&\frac{1}{16\pi}\int d^4x\sqrt{-g}\Big[R-2\Lambda\\&-&2\Big(\nabla_\mu\Phi\nabla^\mu\Phi-2V(\Phi)\Big)-4(\mathscr{L}_{DM}+\mathscr{L}_{I})\Big]\, , \label{action} 
\end{eqnarray}
with $V(\Phi)$ related to the phantom field potential and $\mathscr{L}_{\rm DM}$ and $\mathscr{L}_{I}$ referring to the dark matter Lagrangian density and the interaction between the dark matter and phantom field. Given action the Einstein field equation is written \begin{eqnarray}\label{Eq:gr}
R_{\mu\nu}-\frac{1}{2}g_{\mu\nu}R&+&\Lambda g_{\mu\nu}=8\pi \left(T_{\mu\nu}^{DM}+2\nabla_\mu\Phi\nabla_\nu\Phi\right.\nonumber\\&-&\left. g_{\mu\nu}\nabla_\alpha\Phi\nabla_\alpha\Phi+T_{\mu\nu}^{\rm I}\right)\, ,
\end{eqnarray} 
where $T_{\mu\nu}^{DM}$ is the energy momentum tensor of perfect fluid dark matter profile. Following to Eq.~(\ref{Eq:gr}) one can write the Einstein equations in detailed form \cite{Li-Yang12}  
\begin{eqnarray}
e^{-\delta}\left(\frac{1}{r^2}-\frac{\delta^{\prime}}{r}\right) -\frac{1}{r^2}=\frac{e^{-\delta}}{2}\Phi^{\prime 2}-V(\Phi)+\Lambda-\rho_{\rm DM}\, ,\end{eqnarray}
\begin{eqnarray}
e^{-\delta}\left(\frac{1}{r^2}+\frac{\alpha^{\prime}}{r}\right) -\frac{1}{r^2}=-\frac{1}{2}e^{-\delta}\Phi^{\prime 2}-V(\Phi)-\Lambda\ ,\end{eqnarray}
\begin{eqnarray}
\nonumber
e^{-\delta}\left(\alpha^{\prime\prime}+\frac{\alpha^{\prime 2}}{2}+\frac{\alpha^{\prime}-\delta^{\prime}}{r}-\frac{\alpha^{\prime}\delta^{\prime}}{2}\right)&=&\frac{1}{2}e^{-\delta}\Phi^{\prime 2}\\&-&V (\Phi)-\Lambda\ ,
\end{eqnarray}
where $\alpha$ and $\delta$ are the ansatz which could help find the exact static black hole solutions.   
As a result of above equations,  Schwarzschild-de-Sitter black hole metric surrounded by perfect fluid dark matter field in Boyer-Lindquist coordinates $(t, r, \theta, \varphi)$ can be written as follows:  
\begin{eqnarray}\label{Eq:met1}
ds^2&=&-F(r)dt^2+F(r)^{-1}dr^2 + r^2 d\Omega^2\, ,  
\end{eqnarray}
where 
\begin{eqnarray} 
F(r)=1-\frac{2M}{r}-\frac{\Lambda}{3}r^2+\frac{\lambda}{r} \ln\frac{r}{\vert\lambda \vert}\, .
\end{eqnarray}
Here $M$ is black hole mass and $\Lambda$ and $\lambda$ respectively refer to the cosmological constant and the dark matter field. It is worth noting that in the limit $\Lambda,\lambda \to 0$, the above mentioned metric (\ref{Eq:met1}) takes the the Schwarzschild metric. The stress energy-momentum tensor for the dark matter field distribution yields 
\begin{eqnarray}
\left(T^{\mu}_{\nu}\right)^{DM}={\rm diag}\left(-\rho,p_{r},p_{\theta},p_{\phi}\right)\, ,
\end{eqnarray}
where the density and radial and tangential pressures are given by 
\begin{eqnarray}\label{rho}
\rho=-p_{r}= \frac{\lambda}{8\pi r^3}\,  \mbox{~~~and~~~} p_{\theta}=p_{\phi}=\frac{\lambda}{16\pi r^3}\, .
\end{eqnarray}
Here the point to be noted is that for a dark matter distribution we further focus on the positive value of dark matter profile, i.e $\lambda>0$ which refers to positive energy density. The horizon is located at the positive root of  following equation  
\begin{eqnarray}
r-2M-\frac{\Lambda}{3}r^3+\lambda \ln\frac{r}{\vert\lambda \vert}=0\, ,
\end{eqnarray}
and it solves to give analytical expression for small values of $\Lambda M^2\ll 1$ and $\lambda/M\ll 1$ 
\begin{eqnarray}
r_{h}=M+\sqrt{M^2+\frac{16}{3}M^4\Lambda -2 M \lambda \ln\frac{2M}{\vert\lambda \vert}}\, .
\end{eqnarray}
Note that the horizon radius is given by $r_h=2M$ in case $\Lambda={3\lambda}\ln\frac{2M}{\vert\lambda \vert}/{8 M^3}$. From the above the equation the horizon radius $r_h$ increases as cosmological constant $\Lambda$ is introduced while it decreases once the dark matter field $\lambda$ is taken into account.  

\section{\label{Sec:motion}
Circular orbits around the black hole}

We now investigate particle motion around the Schwarzschild-dS black hole immersed in the perfect fluid dark matter. 
From the Hamilton-Jacobi formalism the Hamiltonian can be taken as constant $H=k/2$ with relation $k=-m^2$, where  $m$ is the mass of test particle. For photons one has to set $k=0$. The action $S$ for Hamilton-Jacobi can be written as  
\begin{eqnarray}\label{separation}
S=-\frac{1}{2}k\tau -Et+L\varphi+S_{r}(r)+S_{\theta}(\theta)\, ,
\end{eqnarray}
where $E$ and $L$ are the usual conserved quantities associated with the time translations and spatial rotations and describe the energy $E$ and angular momentum $L$ of the particle or photon, respectively and  $S_{r}$ and $S_{\theta}$ are functions of only $r$ and $\theta$, respectively. Now it is straightforward to obtain the Hamilton-Jacobi equation in the following form
\begin{eqnarray}
\label{Eq:Sep1}
k=-\frac{E^2}{F(r)}+F(r)
\left(\frac{\partial S_{r}}{\partial r}\right)^2
+\frac{1}{r^2}\left(\frac{\partial S_{\theta}}{\partial \theta}\right)^2+ \frac{L^2}{r^2\sin^2\theta} \, .\nonumber\\
\end{eqnarray}
Due to the spherical symmetry of the spacetime we can restrict the analysis to the equatorial plane $\theta=\pi/2$.  Then from the separability of the action given in Eq.~(\ref{Eq:Sep1}) we obtain the radial equations of motion for particles and photons in the form
\begin{eqnarray} \label{Veff3}
\dot{r}^2={\cal E}^2+F(r)\left(k-\frac{{\cal L}^2}{r^2}\right)={\cal E}^2-V_{\rm eff}(r)\, ,
\end{eqnarray}
where ${\cal E}=E/m$ and ${\cal L}=L/m$ respectively refer to energy and angular momentum per unit mass. For massive particle we have set $k=-1$ so that the effective potential $V_{\rm eff}(r)$ is defined by 
\begin{eqnarray}
V_{\rm eff}(r,{\cal L})&=& F(r)\left(1+\frac{{\cal L}^2}{r^2\sin^2\theta}\right)\, .
\end{eqnarray}
\begin{figure}
 \includegraphics[width=0.45\textwidth]{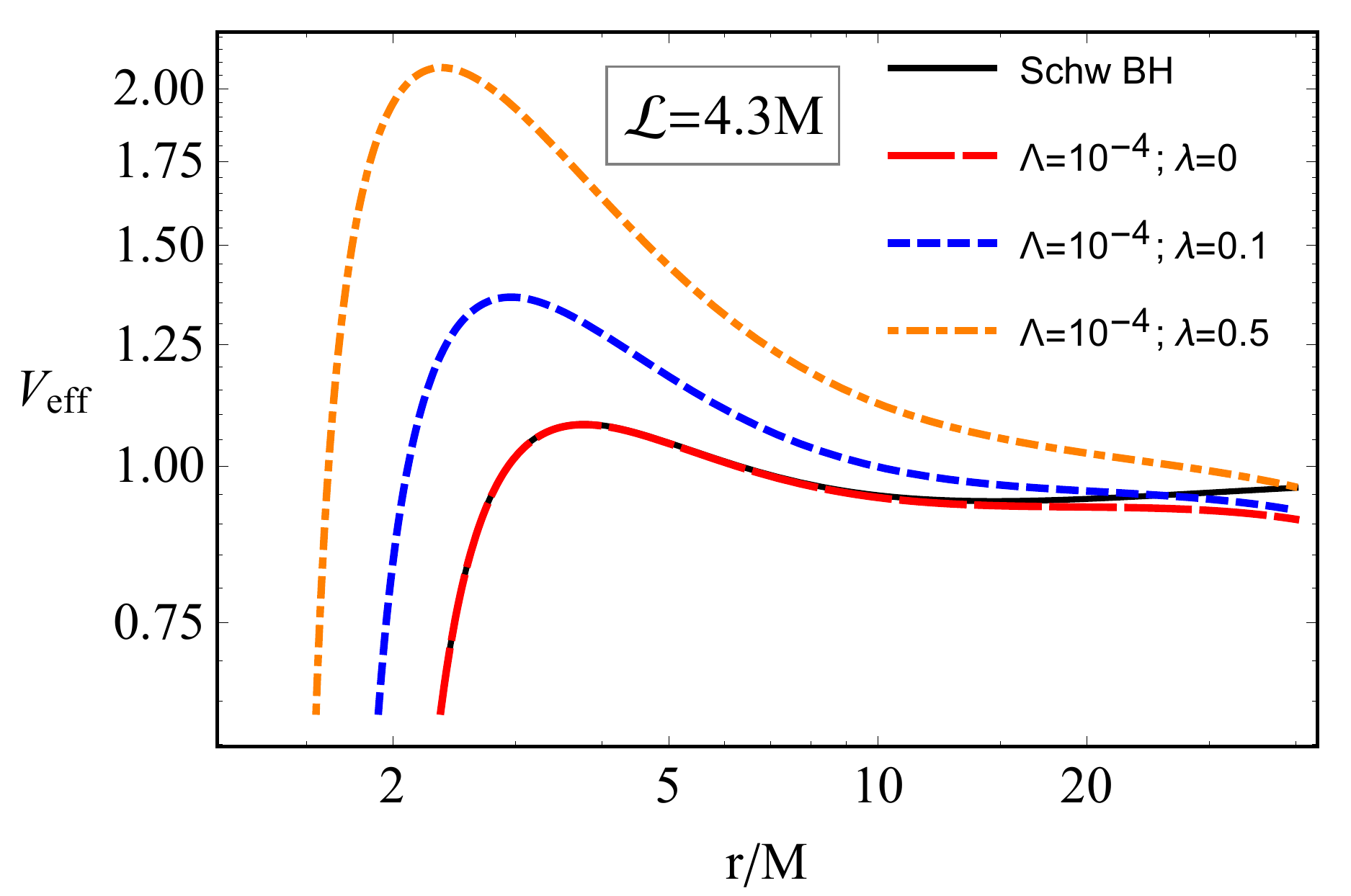}
\caption{ \label{fig:1}Radial profiles of the effective potential for radial motion of test particles around Schwarzschild-de-Sitter BH surrounded by PFDM medium for different values of $\lambda$ and $\Lambda$. Here we use $\Lambda \to \Lambda/M^2$ and $\lambda \to \lambda/M$}\end{figure}

The radial profile of the effective potential is shown in
Fig.~\ref{fig:1} for different values of $\lambda$ for given $\Lambda$. As could be seen from  the radial profile of $V_{\rm eff}$, the curves shift towards left to smaller $r$. In the case of cosmological constant the curves remain at larger $r$.       

\subsection{Stable circular orbits }
\begin{figure}
\centering
\includegraphics[width=0.958\linewidth]{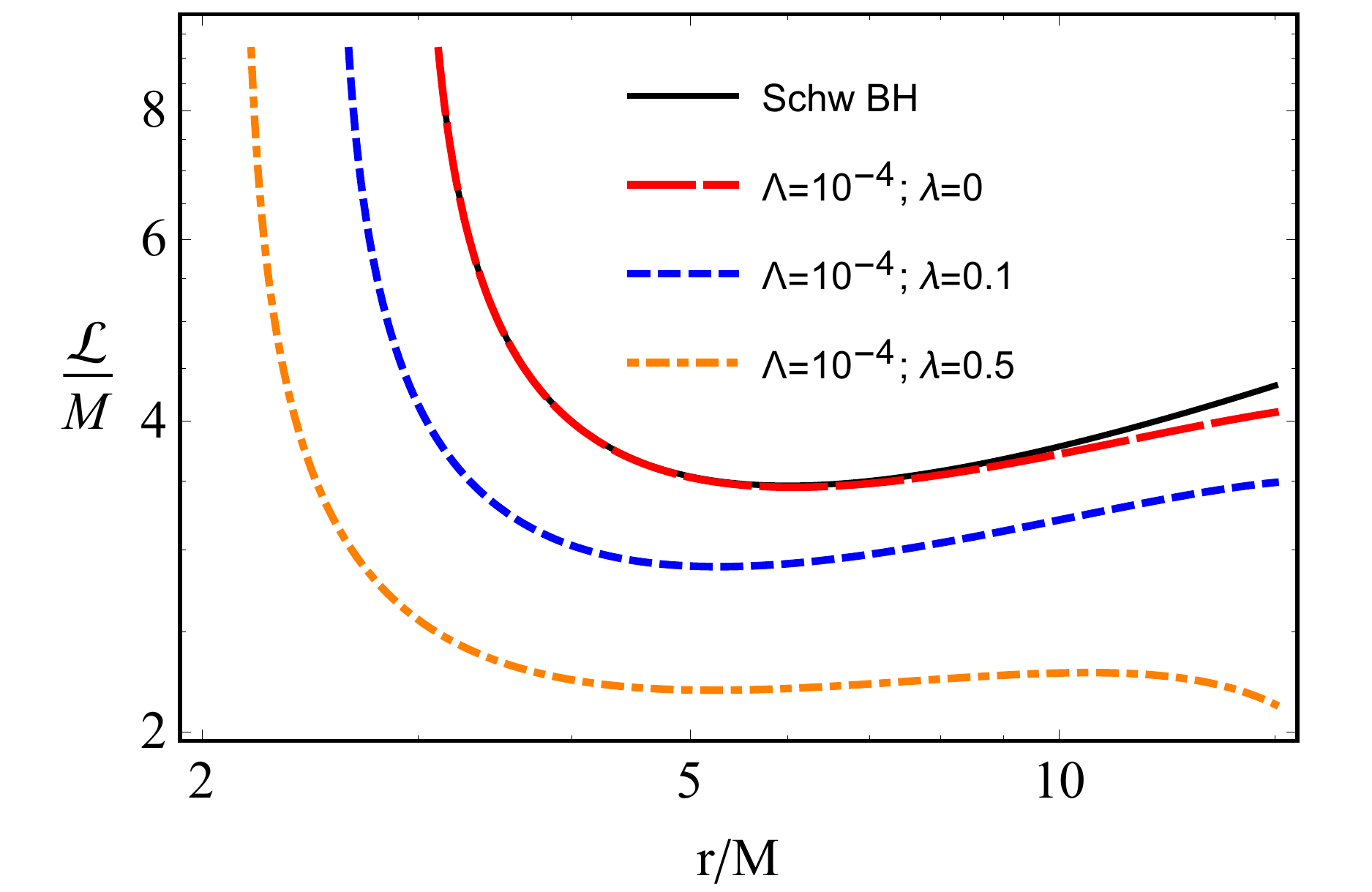}
\includegraphics[width=0.958\linewidth]{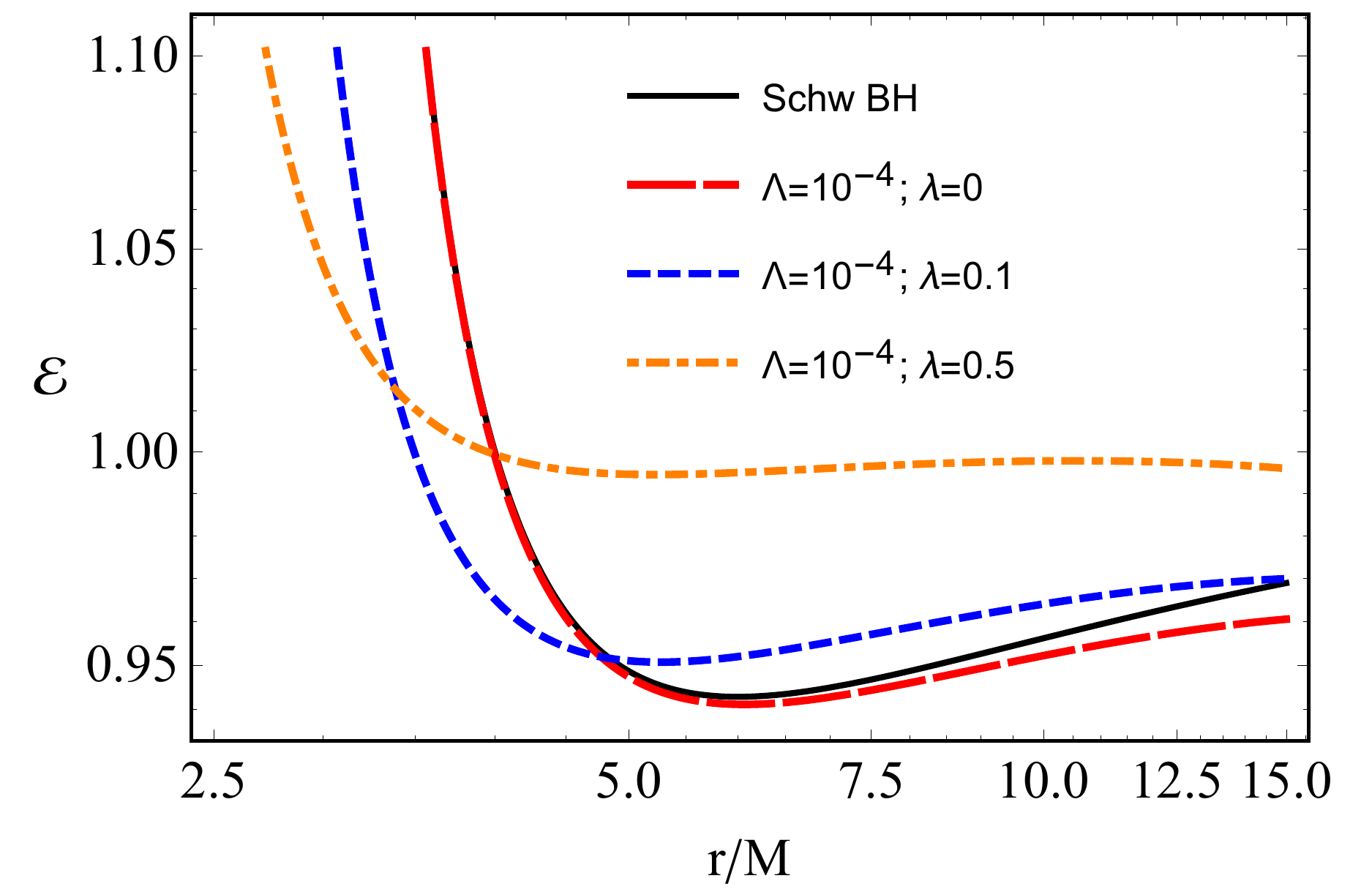}
  \caption{The radial dependence of the specific angular momentum (top panel) and energy  (bottom panel) for test particles orbiting around Schwarzschild-de-Sitter BH surrounded by PFDM for different values of the DM parameter in the case of positive cosmological constant. In this plot we use $\Lambda \to \Lambda/M^2$ and $\lambda \to \lambda/M$. \label{fig:an-en}}\end{figure}

Let us then turn to the the effective potential to find circular orbits for which we should solve simultaneously 
\begin{eqnarray}\label{Eq:condition}
V_{\rm eff}(r,{\cal L})=0, \quad \partial_{r} V_{\rm eff}(r,{\cal L})=0\, .
\end{eqnarray}
The above equation solves to give the radii of circular orbits for given values of ${\cal E}$ and ${\cal L}$.  The radial profiles of the angular momentum ${\cal L}$ and energy ${\cal E}$ at the circular orbits are respectively written in the following form
\begin{eqnarray}\label{Eq:L}
&&{\cal L}^2=\frac{r^2 \left[3 \lambda  \left(1-\ln \frac{r}{\vert\lambda\vert}\right)+6 M-2 \Lambda  r^3\right]}{3 \left[2r-6M-\lambda  \left(1-3 \ln \frac{r}{\vert\lambda\vert}\right)\right]}  \ , \\
&& {\cal E}^2=\frac{2 \left[6 M+\Lambda  r^3-3 r-3 \lambda  \ln \frac{r}{\vert\lambda\vert}\right]^2}{9 r \left[2r-6M-\lambda  \left(1-3 \ln \frac{r}{\vert\lambda\vert}\right)\right]}\ ,
\label{Eq:E} 
\end{eqnarray}
We show respectively the radial profiles of the angular momentum and energy of the circular orbits in Fig.~\ref{fig:an-en}. As $\lambda$ increases for fixed $\Lambda$, the curves of the angular momentum shift towards left to smaller $r$ while they tend to unity from the below for the latter. This means bound circular orbits can exist only for $\lambda <0.5$, i.e. $\cal E \leq $ 1 is always satisfied (see Fig.~\ref{fig:an-en}, bottom panel). As shown in Fig.~\ref{fig:1}, $V_{\rm eff}$ has no minimum and there is a maximum only for $\lambda=0.5$ .    

Let us next consider photon orbits which determine the existence threshold for circular orbits which would exist $r>r_{ph}$.  That is defined by either ${\cal L}^2_{\pm}\rightarrow \infty $ or ${\cal E}^2_{\pm} \rightarrow \infty$, giving the following condition  

\begin{eqnarray}\label{Eq:ph} 
6M-2r+\lambda\left(1-3\ln \frac{r}{\vert\lambda\vert}\right)=0\, ,
\end{eqnarray} 
 which gives explicitly the radii of the photon sphere. From Eq.~(\ref{Eq:ph}) one can easily find the following expression for the photon orbit $r_\text{ph}$ as   
\begin{eqnarray}\label{Eq:ph1}
r_\text{ph}&\approx & 3M+\frac{1}{2} \left(1- \ln \frac{27}{8}\right)\lambda +O(\lambda^2) \, .
\end{eqnarray}
As shown from Eq.~(\ref{Eq:ph1}) this clearly shows that the photon orbit $r_{ph}$ does not depend on cosmological constant $\Lambda$.

Further, we consider the innermost stable circular orbit (ISCO) defined by the minimum value of the angular momentum $\cal L$. In order to have circular orbits the following standard condition should hold well:
\begin{eqnarray}\label{Eq:isco}
\partial_{rr} V_{\rm eff}(r) \geq 0\, .
\end{eqnarray} 
However, to find the ISCO radius one needs to solve $\partial_{rr} V_{\rm eff}(r)=0$ which provides the minimum radius for particle to be in the circular orbit. Thus, we have 
\begin{eqnarray} \label{iscoEq}
\nonumber
0&=&\frac{(3 \lambda +2 r) \left(\lambda -\Lambda  r^3+r\right)}{\lambda  \left(1-3 \ln \frac{r}{\vert\lambda\vert}\right)+6 M-2 r}\\&+&3 \lambda  \left(1-\ln \frac{r}{\vert\lambda\vert}\right)+6 M-5 \Lambda  r^3+r \, .
\end{eqnarray} 
It turns out that it is complicated to solve Eq.~(\ref{iscoEq}) analytically, and thus we explore it numerically. 
\begin{figure}
\centering
   \includegraphics[width=0.45\textwidth]{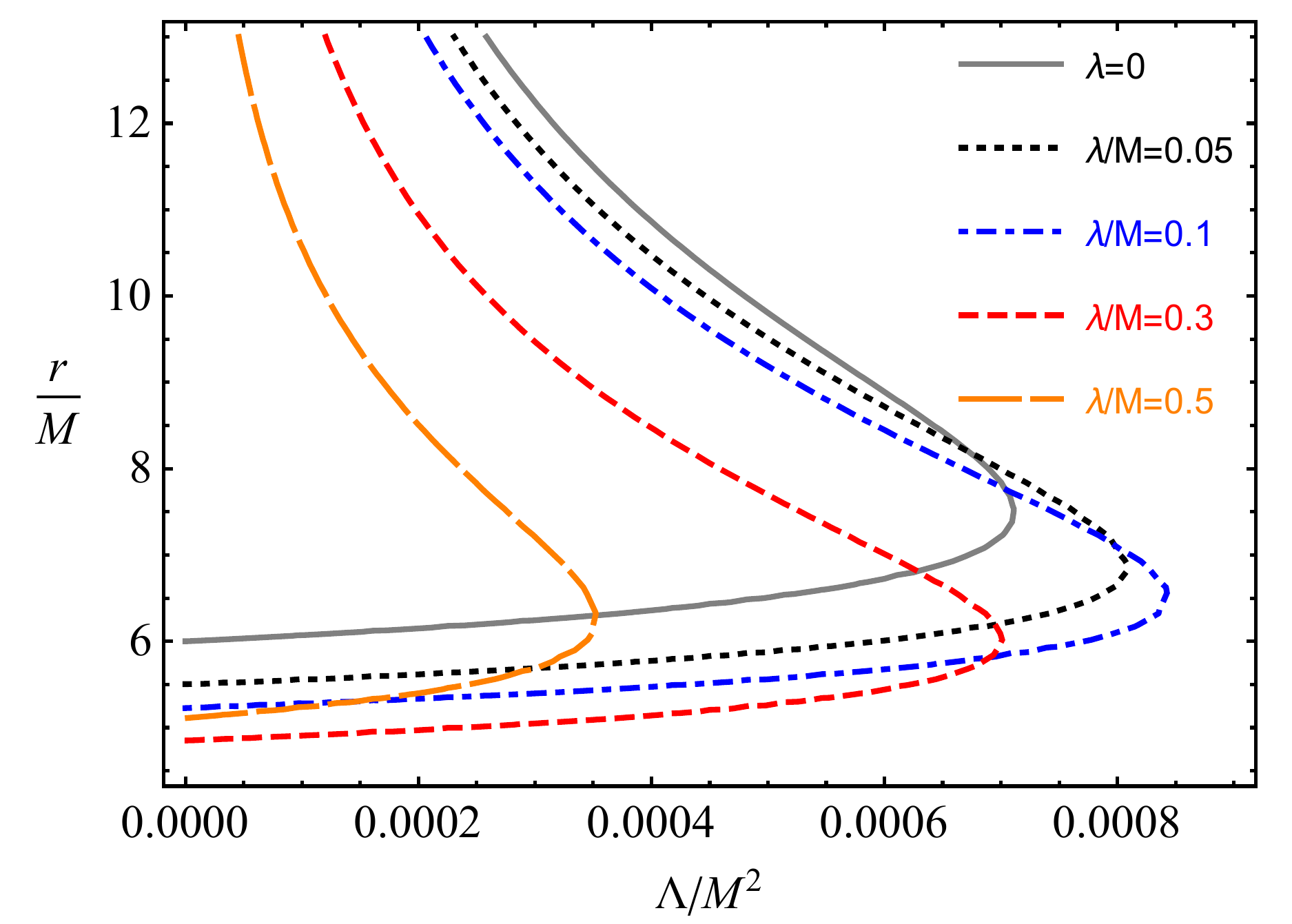}
    \includegraphics[width=0.42\textwidth]{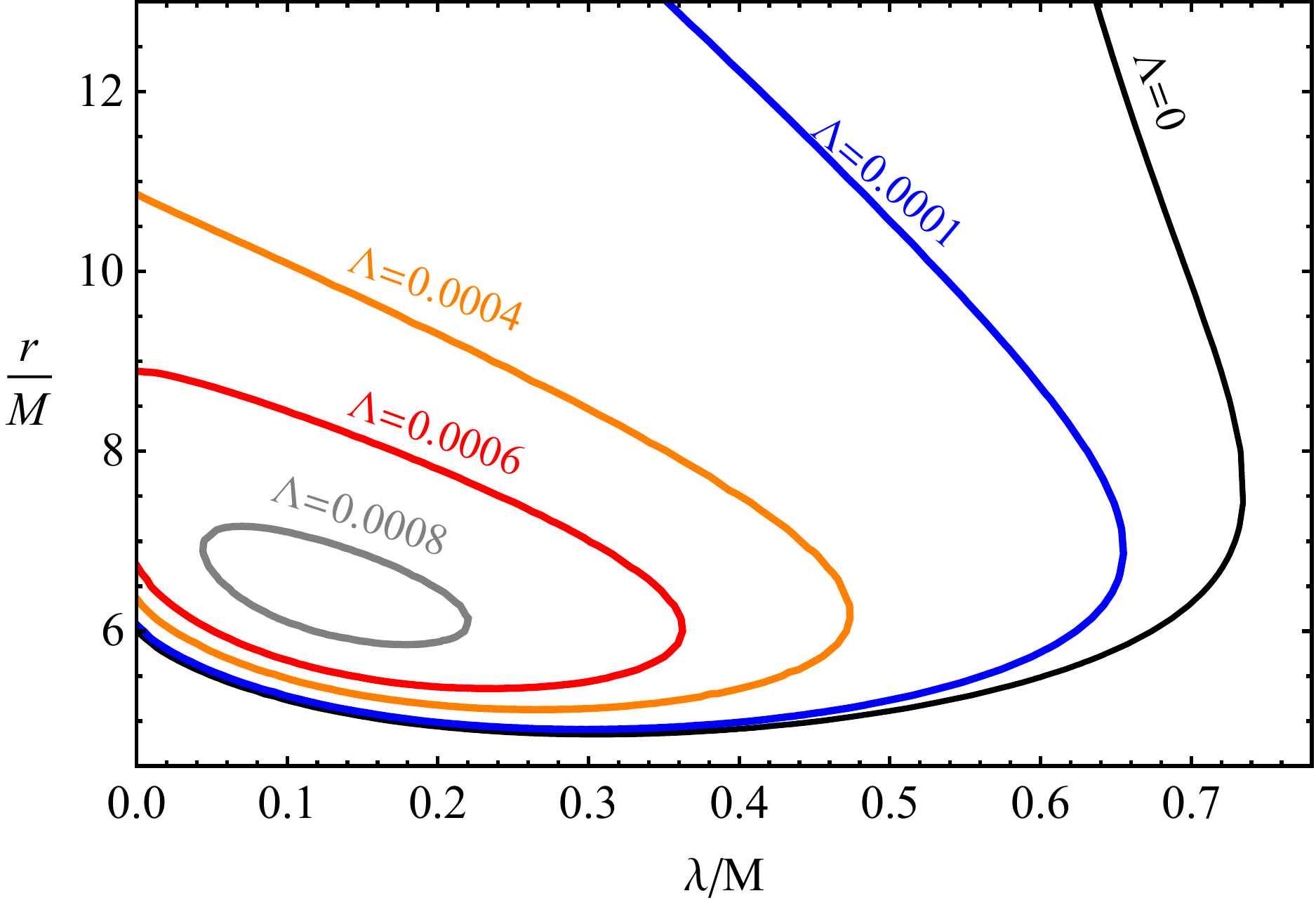}%
 \caption{\label{fig:is_os} Dependence of ISCO and OSCO radius as a function of the cosmological parameter $\Lambda$ in top panel (PFDM parameter $\lambda$ in bottom panel) for different values of PFDM parameter (cosmological constant). The lower (upper) part of the curve determines the ISCO (OSCO). In this plot we use $\Lambda \to \Lambda/M^2$.}\end{figure}

In Fig.~\ref{fig:is_os}, we show the behaviour of the ISCO radius as a function of cosmological constant $\Lambda$ on the top panel and $\lambda$ on the bottom panel. As shown in Fig.~\ref{fig:is_os}, the radius curve consists of two parts, i.e.  the one that grows up from below corresponds to the ISCO and the other that goes down from large $r$ and intersects at the upper limit of the ISCO corresponds to an outermost stable circular orbit (OSCO). Thus, stable circular orbits can only exist between the ISCO from inside and OSCO from outside in the spherically symmetric Schwarzschild-dS black holes vicinity. This, in turn, exhibits striking differences from other spacetime metrics being asymptotically flat at large distances. Also, one can see here that the presence of perfect fluid dark matter or PFDM does play a decisive role between gravitational attraction and cosmological repulsion, thus reducing the values of both the ISCO and OSCO and shrinking the region where circular orbits can exist, see Fig.~\ref{fig:is_os} (top panel). Notice that, for specific lower and upper values of perfect fluid dark matter $\lambda$ in the case of fixed $\Lambda=0.0008$ the particles can have only one circular orbit as the ISCO and OSCO coincide, see Fig.~\ref{fig:is_os} (bottom panel). This happens because the gravitational attraction due to the contribution of the perfect fluid dark matter dominates over cosmological repulsion. This gives rise to a remarkable result that there exist double matching values of the ISCO and OSCO in the Schwarzschild-dS spacetime in the presence of dark matter field in contrast to the the Schwarzschild-dS metric. It is worth noting that the ISCO radius decreases for small values of $\lambda$. It does however increase for larger values of $\lambda$. This happens because for larger values of $\lambda$ its effect can turn repulsive due to the repulsive nature of the radial pressures. Thus, there exist no stable circular orbits anymore for larger values of $\lambda$ at large $r$, as seen in Fig.~\ref{fig:is_os} (bottom panel).   

\begin{figure}
\centering
  \includegraphics[width=0.45\textwidth]{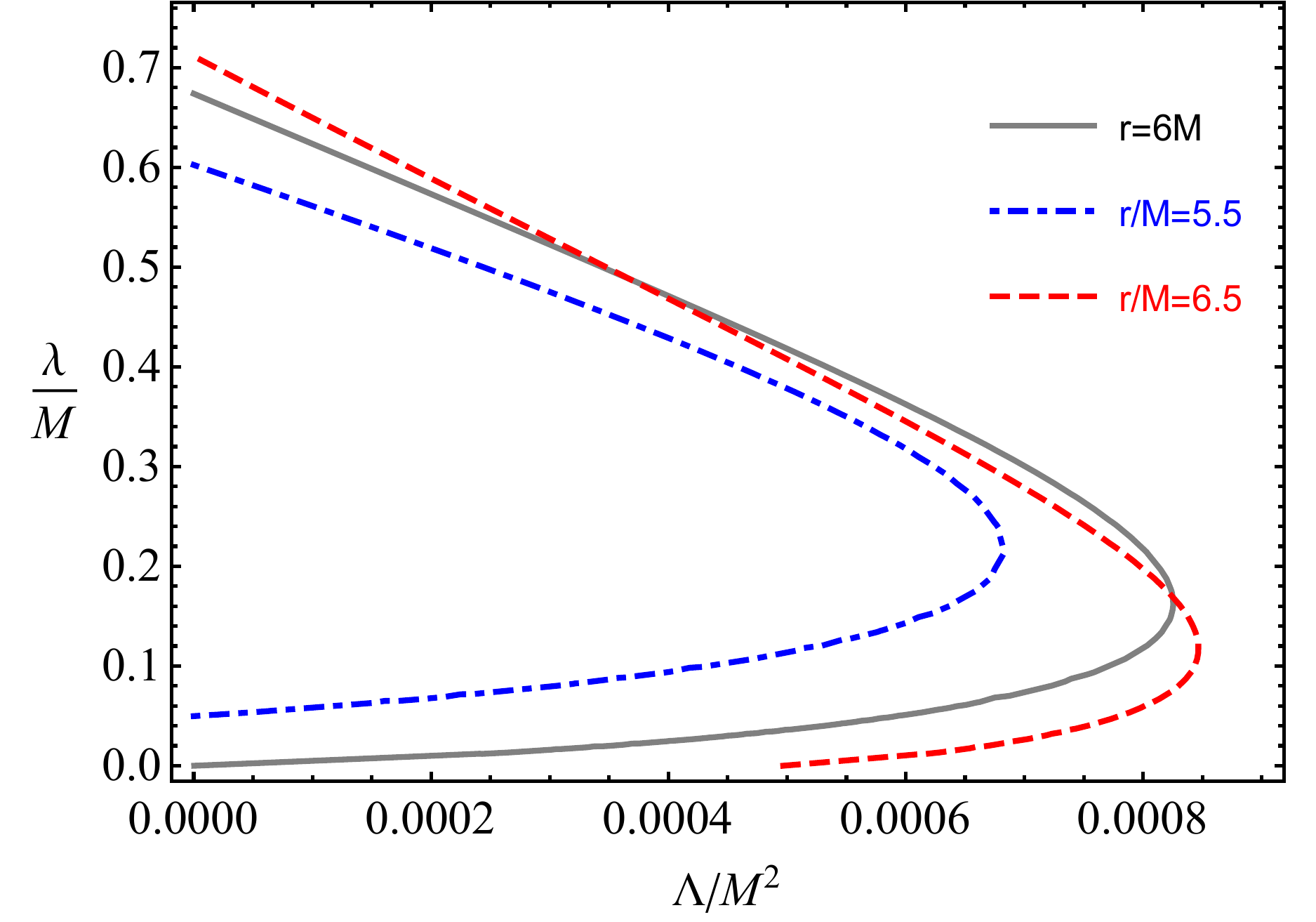}
\caption{\label{fig:dm_cc} The dependence of the cosmological constant $\Lambda$ on the perfect fluid dark matter parameter $\lambda$ for a test particle on circular orbit for which the ISCO radius is the same as the ISCO in the Schwarzschild spacetime.}
\end{figure}

Next, let us find the values of $\lambda$ and $\Lambda$ for which one can compare the behaviour of test particles with the one in the Schwarzschild case.  In Fig.~\ref{fig:dm_cc}, we demonstrate the relation between $\Lambda$ and $\lambda$ for which a test particle can have the same orbits as particles on circular orbits around the Schwarzschild black hole. 
Or, in other words, the combined effects of cosmological constant $\Lambda$ and dark matter $\lambda$ can mimic the Schwarzschild black hole, thus providing the same ISCO radius, as seen in Fig.~\ref{fig:dm_cc}. 

\subsection{Schwarzschild de-Sitter BH surrounded by PFDM versus Kerr BH:  a comparison of the ISCO radius}

In fact, the existence of perfect fluid dark matter parameter and black hole spin shrinks the size of the ISCO radius. Since they exhibit a repulsive gravitational charge, it would be difficult for a far away observer to distinguish whether black hole has spin or is immersed in perfect fluid dark matter field.
With this motivation,  we further explore the degeneracy values of perfect fluid dark matter parameter of Schwarzschild dS black hole and the spin parameter of Kerr BH for which they have the same ISCO radius. 
Thus, in astronomical observations, distinguishing black hole spacetimes with direct or indirect way still remains the challenging question.    

Let us here write the ISCO radius for test particles orbiting around rotating Kerr BHs for retrograde and prograde orbits~\cite{Bardeen72}
\begin{eqnarray}
r_{\rm isco}= 3 + Z_{2} \pm \sqrt{(3- Z_{1})(3+ Z_{1} +2 Z_{2} )} \ ,
\end{eqnarray}
with
\begin{eqnarray} \nonumber
Z_1 &  = & 
1+\left( \sqrt[3]{1+a}+ \sqrt[3]{1-a} \right) 
\sqrt[3]{1-a^2} \ ,
\\ \nonumber
Z_{2}^2 & = & 3 a^2 + Z_{1}^2 \ .
\end{eqnarray}

We now turn to the main motivation to determine values of spin parameter $a$ and perfect fluid dark matter parameter $\lambda$ for which the ISCO radius takes the same values. 
\begin{figure} \centering  \includegraphics[width=0.9\linewidth]{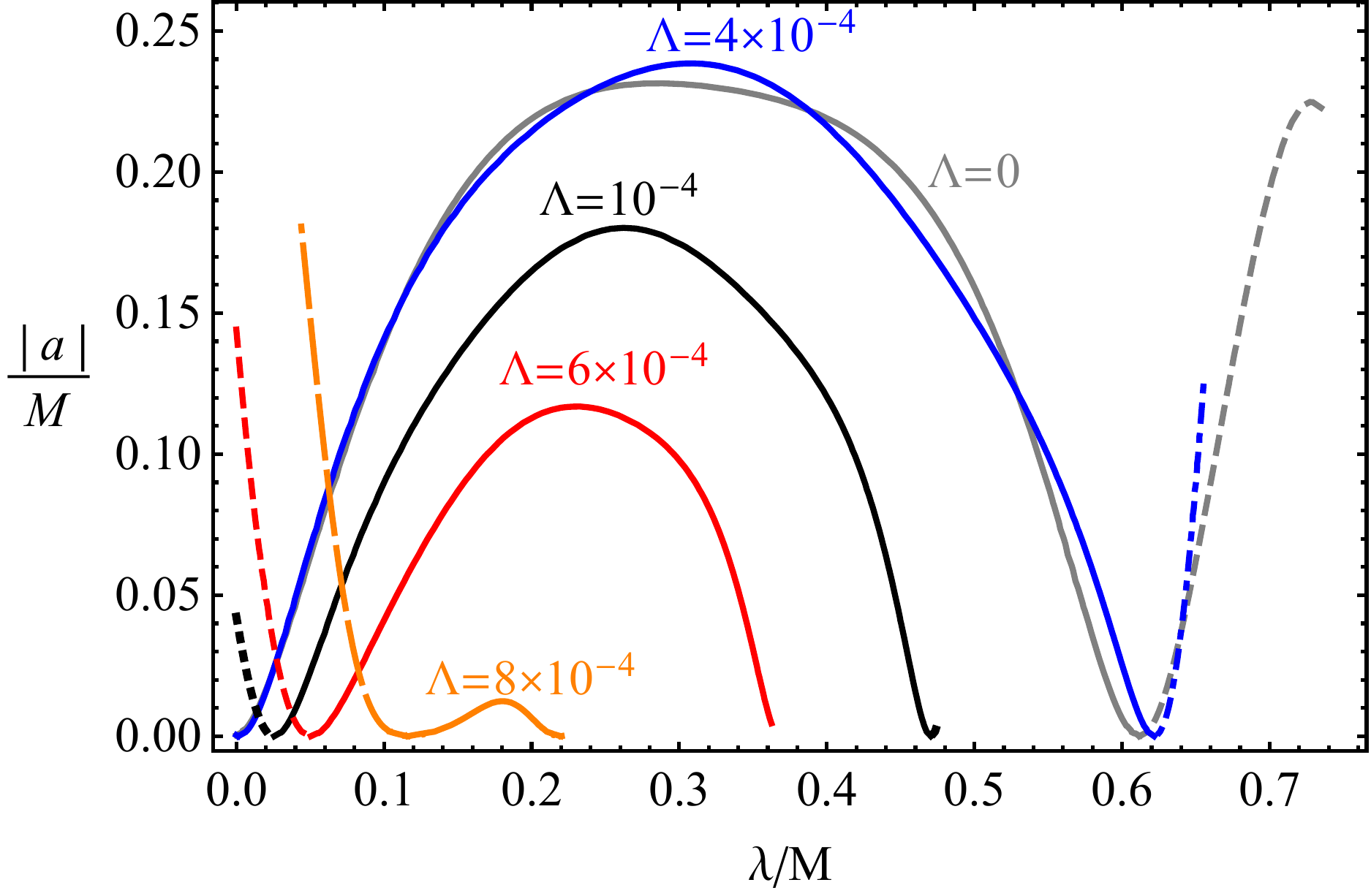}
\caption{
The degeneracy relation between the values of the spin parameter of Kerr BH and the PFDM parameter $\lambda$ for which the radius of ISCO for Kerr BH is the same as the ISCO for the Schwarzschild BH surrounded by PFDM in de-Sitter spacetime. In this plot we use $\Lambda \to \Lambda/M^2$. \label{mimic1}}\end{figure} 

In Fig.~\ref{mimic1} we show the degeneracy between spin parameter $a$ and perfect fluid dark matter parameter $\lambda$ for the location of the ISCO for different values of cosmological constant $\Lambda$. As shown in Fig.~\ref{mimic1}, this clearly shows that an increase in the value of $\Lambda$ leads to mimic smaller values of the spin parameter $a$ for given values of $\lambda$. This happens because the cosmological constant is interpreted as an attractive gravitational charge.

\subsection{Stability of orbits }

In this subsection, we study the stability (instability) of circular orbits for test particles orbiting around a Schwarzschild dS black hole surrounded by perfect fluid dark matter. For that we consider Lyapunov exponent (LE) which represents a measure of the average rate, by which nearby trajectories can converge or diverge in the phase space~\cite{Li2019EPJP}.  For a geodesic stability analysis in terms of Lyapunov exponents (by $\lambda_{L}$ we would mean LE) one needs to deal with the following equation ~\cite{Sharif2017EPJC,Cardoso2009PhRvD}
  \begin{eqnarray}\label{Eq:LE}
  &&\lambda_{L}(r;{\cal L},\lambda,\Lambda)=\sqrt{\frac{-\partial_{rr}V_{\rm eff}(r;{\cal L},\lambda,\Lambda)}{2\dot{t}^2}}\, .
   \end{eqnarray}
Here LE would play an important role as the main mathematical tool in studying the asymptomatic behaviour of trajectories (i.e. the behavior of the orbits). Accordingly, for test particle trajectories to be on stable orbits $\lambda_{L}<0$ always. For $\lambda_{L}=0$ bifurcation points occur, while for $\lambda_{L}>0$ particle trajectory becomes more chaotic. By imposing these three conditions we analyse stability (instability) of particle orbits.
Here, recalling Eq.~\ref{Eq:LE} we derive the following equation for for stable circular orbits
\begin{eqnarray}
\nonumber
\lambda_{L} &=&\frac{3 \lambda  \ln \frac{r}{\vert \lambda \vert }-6 M-\Lambda  r^3+3 r}{\sqrt{6} r^2 \left(6 M+\Lambda  r^3-3 r-3 \lambda \ln \frac{r}{\vert \lambda \vert }\right)}\\\nonumber
&\times & \Bigg\{2 \Lambda  r^3 (4 \lambda +15 M-4 r)-3 \lambda  \ln \frac{r}{\vert \lambda \vert } \\ \nonumber &\times & \left(3 \lambda  \ln \frac{r}{\vert \lambda \vert }-4 \lambda -12 M+5 \Lambda  r^3+r\right)\\ &-& 6 \left(\lambda ^2+6 M^2+4 \lambda  M-M r\right)\Bigg\}^{\frac{1}{2}}
\end{eqnarray}
 \begin{figure}\centering \includegraphics[width=0.85\linewidth]{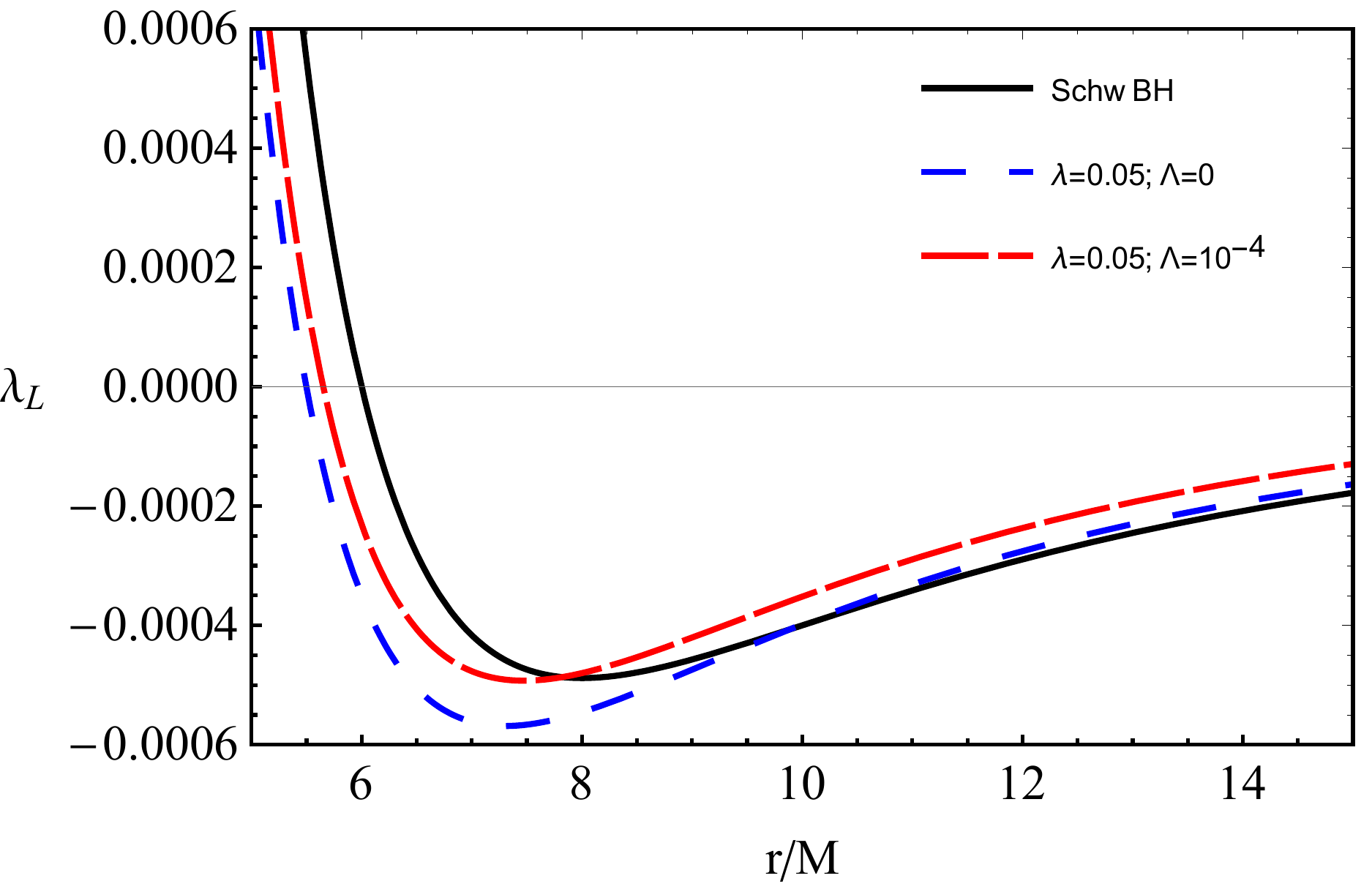} 
  \caption{The radial dependence of LE for the motion of test particles around Schwarzschild black hole surrounded by PFDM medium in de-Sitter spacetime for the different values of the cosmological constant $\Lambda$ and the parameter $\lambda$. In this plot we use $\Lambda \to \Lambda/M^2$ and $\lambda \to \lambda/M$.  \label{Lyavsrbeta}}\end{figure} 
    
In Fig.~\ref{Lyavsrbeta}, we show the radial dependence of the Lyapunov exponent for test particle around the Schwarzschild-dS black hole immersed in perfect fluid dark matter field. As can be seen from Fig.~\ref{Lyavsrbeta} the region where stable circular orbits can exist expands with increasing the value of dark matter $\lambda$ in the case of fixed $\Lambda$. For small radii particle trajectory starts becoming more chaotic as there exist no stable circular orbits for test particles. One can also see that bifurcation points occur when the curves touch the horizontal axis, i.e. $\lambda_{L}=0$ which determines the location of the ISCO. As shown in Fig.~\ref{Lyavsrbeta} the ISCO radius shifts towards left to smaller $r$ as a consequence of the increase of perfect fluid dark matter $\lambda$.

\subsection{The energy efficiency}

The Keplerian accretion around an astrophysical BH has been modelled by Novikov and Thorne \cite{Novikov73} that thin disks geometrically reflect the effects of spacetime properties on circular geodesics. Here we define the energy efficiency of the accretion disk around a black hole. That is the highest energy can be extracted by the accretion disk due to the falling in matter into the black hole.  The expression for energy efficiency is defined by \cite{Bardeen73}
\begin{equation}
\eta=1-{\cal E}_{\rm ISCO}\, ,
\end{equation}
with test particle energy ${\cal E}_{\rm ISCO}$ at the ISCO. For that we use the energy expression Eq.(\ref{Eq:E}) for particles at the ISCO by solving Eq. (\ref{Eq:isco}). for For an exact analytical form one needs to solve these two equations simultaneously, but it turns out to be very complicated. Thus, we shall explore the combined effects of $\lambda$ and $\Lambda$ on the energy efficiency numerically and provide plots for details.  
\begin{figure} \centering \includegraphics[width=0.98\linewidth]{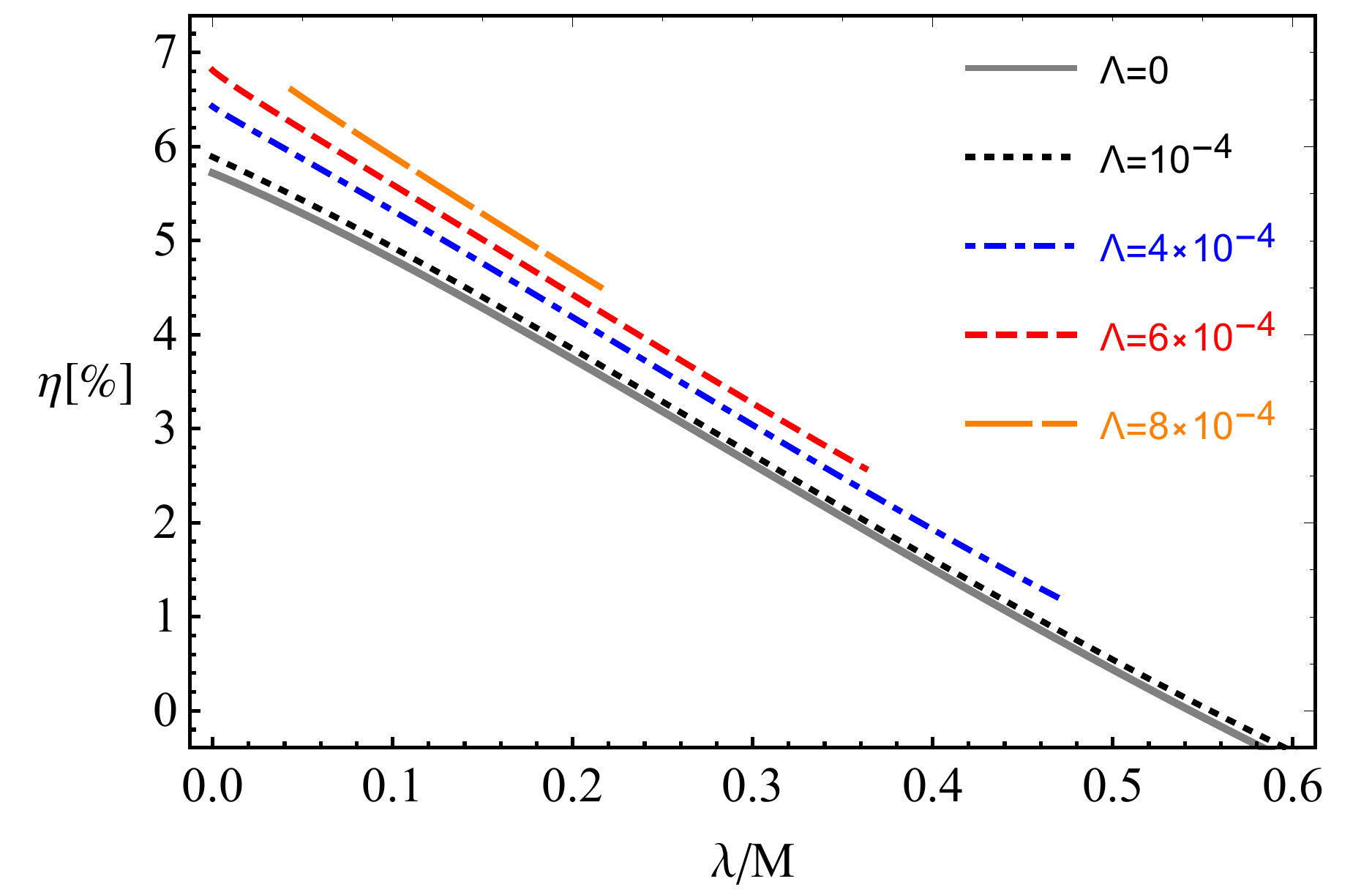} 
\caption{Efficiency of energy release from accretion disc around Schwarzschild BH surrounded by PFDM in de-Sitter as a fuction of the parameter $\lambda$ for the different values of $\Lambda$. In this plot we use $\Lambda \to \Lambda/M^2$. \label{efficiency}} \end{figure}

In Fig.~\ref{efficiency} we show the dependence of energy efficiency released by falling in particle (matter) on the perfect fluid dark matter parameter $\lambda$ for different values of the cosmological constant $\Lambda$. One can see that as a consequence of the increase of $\lambda$ the energy efficiency decreases while it slightly increases as $\Lambda$ increases. 

It is well known that bolometric luminosity of the black hole accretion disk is defined by the following relation \cite{BokhariPhysRevD2020}
\begin{eqnarray}
L_{bol}=\eta \dot{M}c^2\, ,
\end{eqnarray}
with $\dot{M}$ being the rate of accretion matter falling into the black hole. From an astrophysical point of view, there are important issues in observations of bolometric luminosity, one of which is related to the determination of the black hole type. Since theoretical analysis/models, proposed to measure the bolometric luminosity, show similar characters for effects of parameters in any two different gravity models, we turn to the question about whether the combined effects of perfect fluid dark matter and cosmological constant can mimic the effects of bolometric luminosity and the spin parameter of Kerr black hole, thus providing the same value of the energy efficiency.

The expression of the energy efficiency released from accretion disk of Kerr black hole is defined by \cite{Bardeen72}
\begin{eqnarray}
{\cal E}_{\rm Kerr}(r,a)=\frac{a+r^{3/2}-2 \sqrt{r}}{r^{3/4} \sqrt{2 a+r^{3/2}-3 \sqrt{r}}}
\end{eqnarray}

\begin{figure}  \centering
\includegraphics[width=0.93\linewidth]{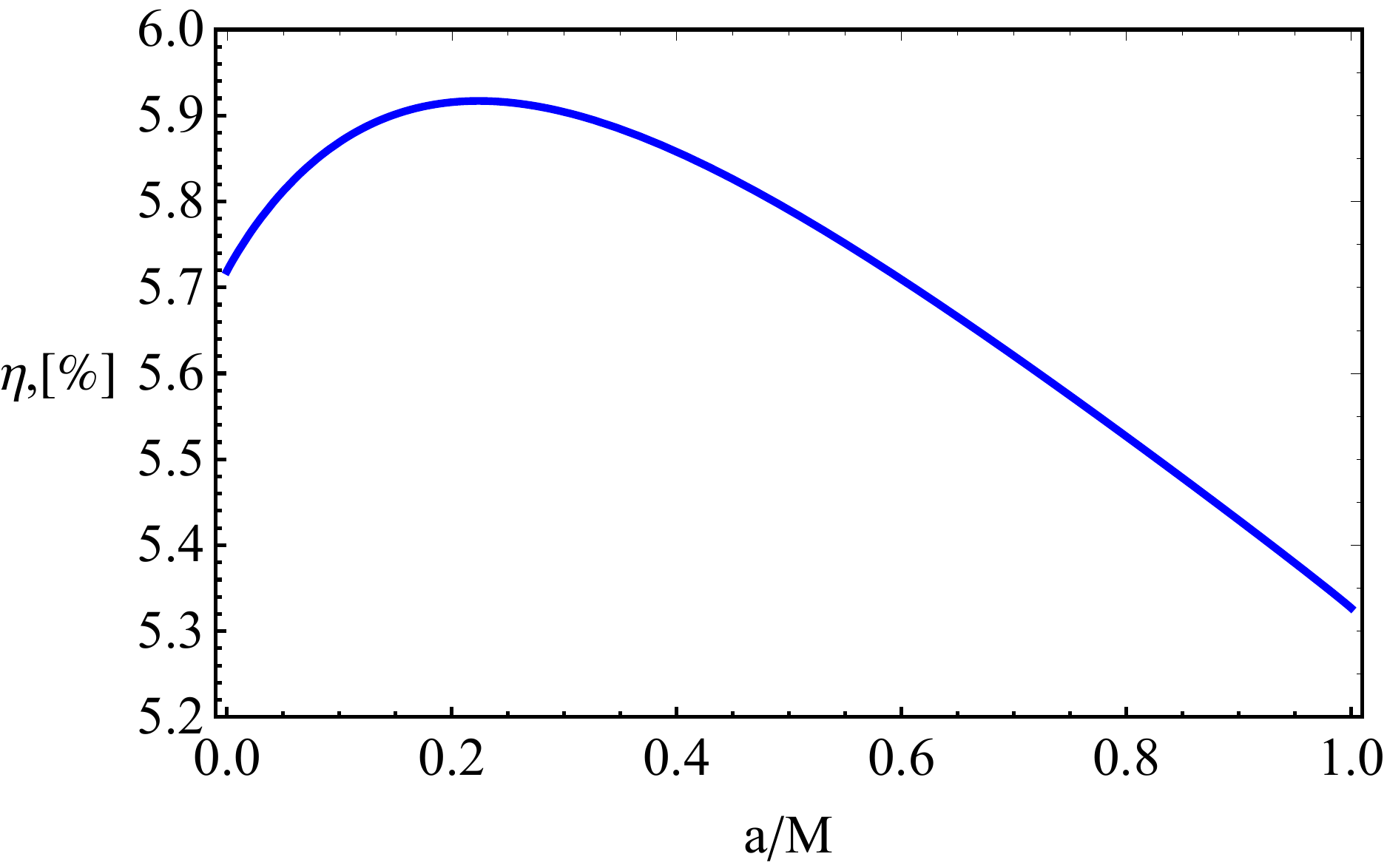}
\includegraphics[width=0.90\linewidth]{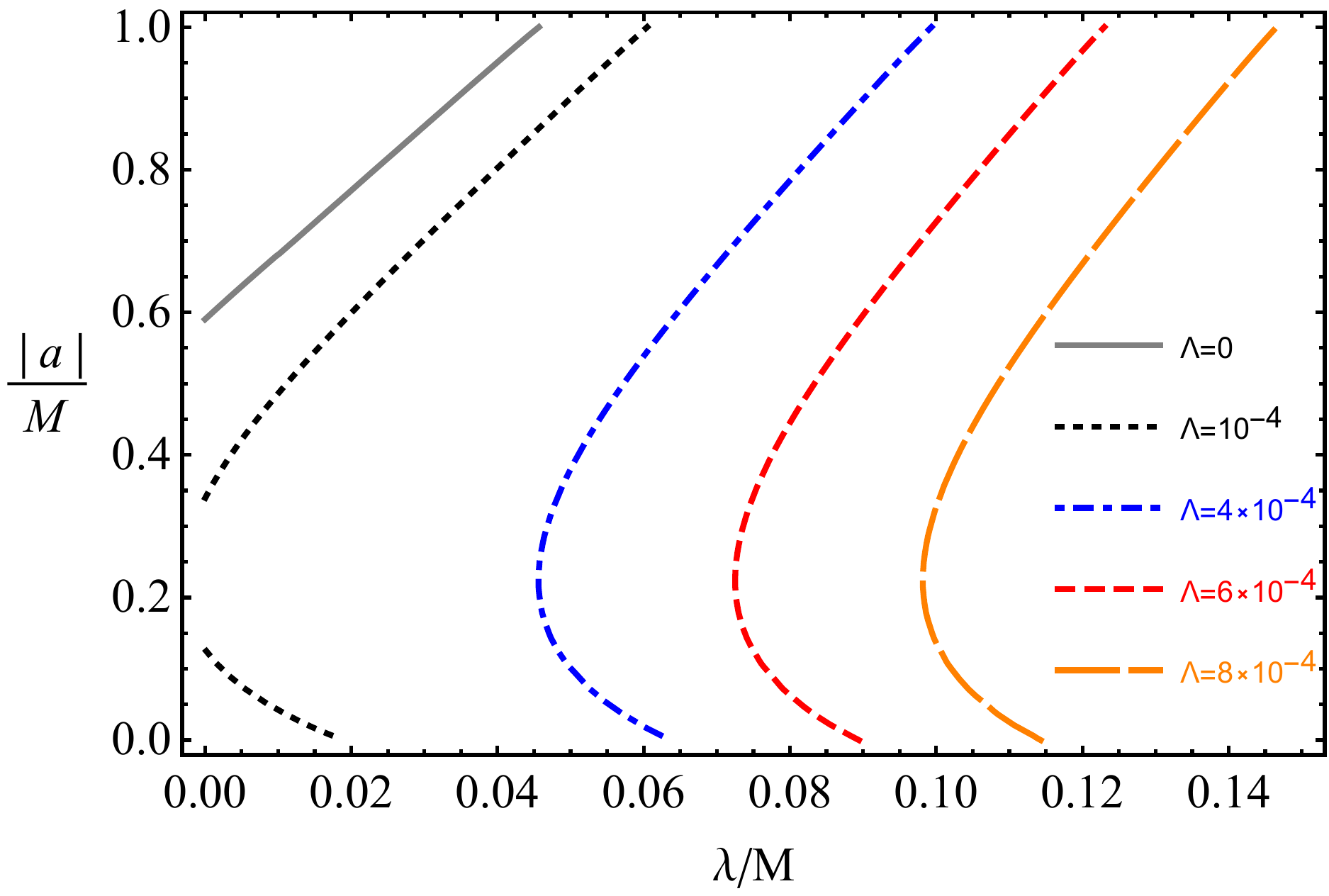}	
\caption{Dependence of the energy release efficiency from Kerr BH's retrograde accretion disk from its spin parameter (top panel) and degeneracy relations between spin of the Kerr BH and the DM parameter providing the same energy efficiency in case when the matter rate down flow to central BH is equal in the both cases.  In this plot we use $\Lambda \to \Lambda/M^2$. \label{effQvsa}} \end{figure}

Fig.~\ref{effQvsa} shows the relation between the black hole spin parameter and perfect fluid dark matter for which the energy efficiency released from retrograde orbit onto accretion disk in the Kerr black hole geometry is the same as the energy efficiency in the Schwarzschild-dS black hole geometry in the perfect fluid dark matter for different values of cosmological constant. 
{As can be seen from the top panel, two different values of the spin parameter can provide the same energy efficiency $\eta\,[\% ]$ given in the range $(5.7191,5.9169)$.}
We also show the degeneracy for spin parameter which mimics entirely the effect of dark matter parameter, see in Fig~\ref{effQvsa} (bottom panel). As shown on the bottom panel, this clearly shows that the degeneracy relations shift towards right to larger $\lambda$ due to the inclusion of the effect of cosmological constant.

\section{Fundamental frequencies}\label{Sec:frequency}

One of the highly energetic phenomenon in the Universe is an astrophysical system comprising an accretion disk surrounding a compact star or a black hole. Occasionally the compact source-accretion disk system emits a jet along with energetic x- and $\gamma-$ rays. Such astrophysical systems are usually termed microquasars which exihibit  quasiperiodic oscillations or QPOs and characterized by high and low frequency peaks in their power spectrum. So far there is no deep understanding as to how the QPOs are produced but simplified models are based on the assumption of the resonance of orbital and radial oscillation frequencies of test neutral particles orbiting the compact source in the perturbed circular orbits. We are mainly interested in the high frequency QPOs which are modelled as Kerr black holes with charged or neutral particles in the perturbed ISCOs producing epicyclic frequencies. 

In this section, we study fundamental frequencies of test particles around stable circular orbits in the background geometry of Schwarzschild-dS black hole immersed in perfect fluid dark matter, i.e. Keplerian (orbital) frequency, frequencies of radial and latitudinal oscillations with their applications to upper and lower frequencies of twin peak QPO objects for various models. 

\subsection{Keplerian frequency}

The angular velocity of particle orbiting a BH's accretion disk measured by an observer located at infinity, known as the Keplerian frequency, is defined by 
\begin{eqnarray}\label{Omega}
\Omega_{K}=\frac{d\phi}{dt}=\frac{\dot{\phi}}{\dot{t}}\ , 
\end{eqnarray}
Following Eq.(\ref{Omega}) the Keplerian frequency of a test particle for a static black hole spacetime reads as
\begin{eqnarray}\label{omega}
\Omega_K=\sqrt{\frac{f^{\prime}(r)}{2r}}\, ,    
\end{eqnarray}
while for Schwarzschild-dS black hole with perfect fluid dark matter field Eq.(\ref{Omega}) yields 
\begin{eqnarray}\label{omega}
\Omega_K=\frac{\sqrt{M- \frac{\Lambda}{3}  r^3+\frac{\lambda}{2}  \left[1-\ln \frac{r}{\vert \lambda \vert }\right] }}{r^{3/2}}\ .   
\end{eqnarray}

Our aim here is to understand the results of fundamental frequencies on the basis of analysis, thus we shall define them in the standard unit (i.e. Hz) 
\begin{eqnarray}
\nu = \frac{1}{2\pi}\frac{c^3}{GM} \Omega\ , [{\rm Hz}]\ .
\end{eqnarray}
For that we convert geometrical unit (i.e. $\rm 1/cm^{1/2}$) into the  standard one (Hz) for which we use $c=3\cdot10^{10} \rm cm/sec$ and $G=6.67\cdot 10^{-8} \rm cm^3/(g\cdot sec^2)$.
\begin{figure} \centering \includegraphics[width=0.98\linewidth]{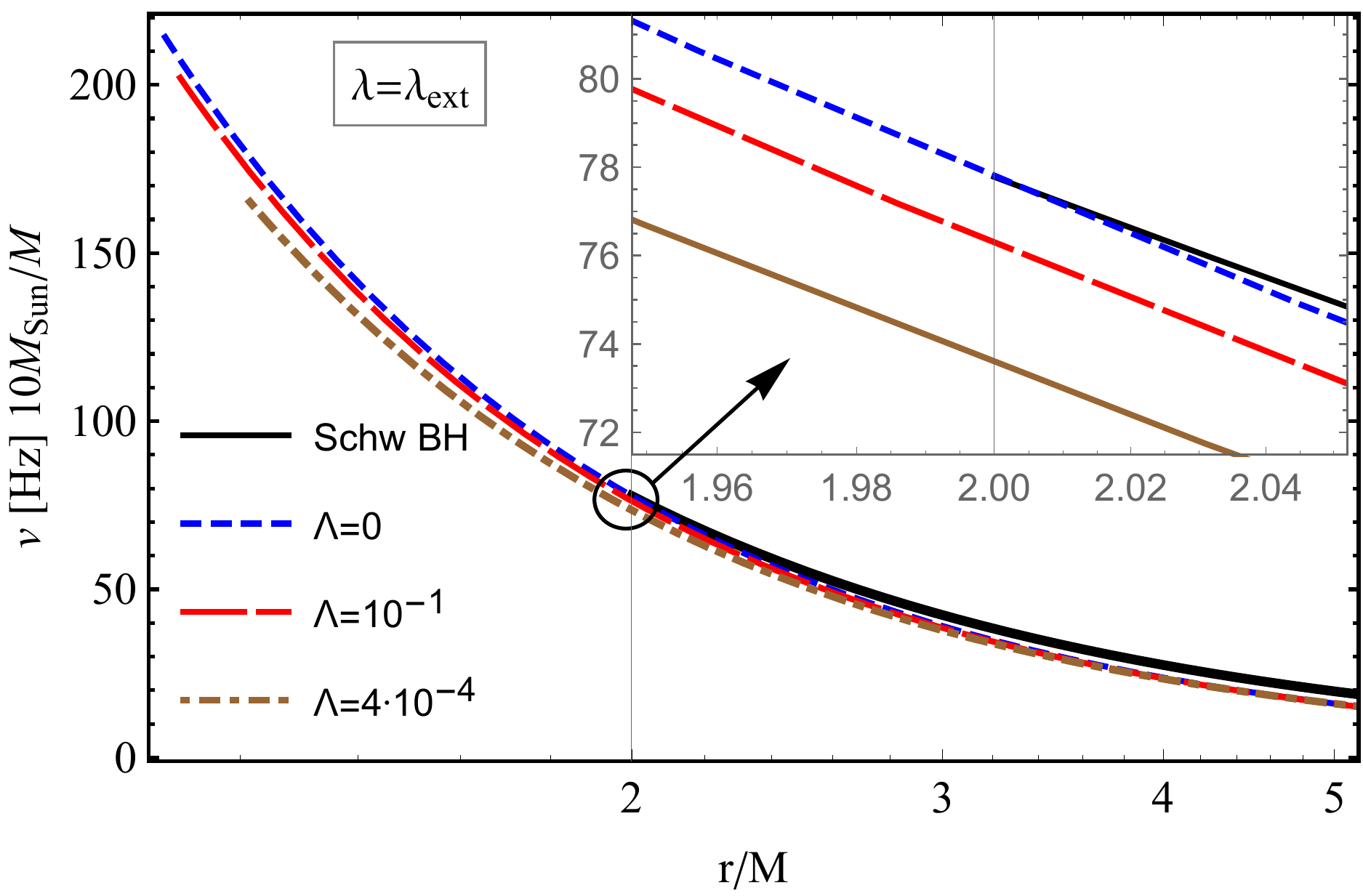}	\caption{Radial dependence of Keplerian frequencies of test particles around a Schwarzschild-de-Sitter BHs for the different values of the cosmological constant. In this plot we use $\Lambda \to \Lambda/M^2$ and $\lambda \to \lambda/M$. \label{Keplerian}} \end{figure}

The radial dependence of the Keplerian frequencies for test particles around Schwarzschild-dS black hole is shown in Fig.~\ref{Keplerian}. From Fig.~\ref{Keplerian}, one can see that as a consequence of the combined effects of cosmological constant and dark matter the Keplerian frequency gets slightly decreased.

\subsection{Harmonic oscillations}

 We consider here test particle motion on the stable circular orbits. For a small perturbation $r\to r_0+\delta r$ and $\theta\to \theta_0+\delta\theta$ the particle oscillates with radial and latitudinal fundamental frequencies with the so-called epicyclic motion. Since test particle is slightly perturbed around $r_0$ one can expand the effective potential in small $\delta r$ and $\delta\theta$ as follows: 
\begin{eqnarray}\label{Vexpand}
\nonumber
&&V_{\rm eff}(r,\theta)=V_{\rm eff}(r_0,\theta_0)
\\ \nonumber
&&+\delta r\,\partial_r V_{\rm eff}(r,\theta)\Big|_{r_0,\theta_0} +\delta\theta\, \partial_\theta V_{\rm eff}(r,\theta)\Big|_{r_0,\theta_0}
\\\nonumber
&&+\frac{1}{2}\delta r^2\,\partial_r^2 V_{\rm eff}(r,\theta)\Big|_{r_0,\theta_0}+\frac{1}{2}\delta\theta^2\,\partial_\theta^2 V_{\rm eff}(r,\theta)\Big|_{r_0,\theta_0}
\\
&&+\delta r\,\delta\theta\,\partial_r\partial_\theta V_{\rm eff}(r,\theta)\Big|_{r_0,\theta_0}+{\cal O}\left(\delta r^3,\delta\theta^3\right)\ .
\end{eqnarray}
From the above equation we eliminate the first three terms by imposing the condition Eq.~(\ref{Eq:condition}) for stable circular orbits and then keep only the second order terms. 
We then substitute Eq.~(\ref{Vexpand}) into Eq.~(\ref{Veff3}), so that we obtain harmonic oscillator equations for the radial and latitudinal oscillations in the following form
\begin{eqnarray}
\delta \ddot{r}+\Omega_r^2 \delta r=0\, \qquad \mbox{and} \qquad  \delta\ddot{\theta}+\Omega_\theta^2 \delta\theta=0\, ,   
\end{eqnarray}
where $\Omega_r$ and $\Omega_\theta$, which respectively refer to the radial and latitudinal angular frequencies measured by a distant observer, are defined as
\begin{eqnarray}
&&\Omega_r^2=-\frac{1}{2g_{rr}}\partial_r^2V_{\rm eff}(r,\theta)\Big|_{\theta=\pi/2}\ ,
\\
&&\Omega_\theta^2=-\frac{1}{2g_{\theta\theta}}\partial_{\theta^2} V_{\rm eff}(r,\theta)\Big|_{\theta=\pi/2}\, .
\end{eqnarray}
The radial and latitudinal frequencies are then given by 
\begin{eqnarray}\label{wr}
\nonumber
\Omega_r &=&\Omega_K \Bigg\{1-\frac{6 M}{r}+3 \Lambda  r^2-\frac{\lambda  }{r}\left(1-3 \ln \frac{r}{\vert \lambda \vert }\right) \\ &-&\frac{3 \left(1-\Lambda  r^2+\frac{\lambda }{r}\right) \left(2 \Lambda  r^2+\frac{\lambda }{r}\right)}{\frac{3 \lambda }{r}\left(1-\ln \frac{r}{\vert \lambda \vert}\right)+\frac{6 M}{r}-2 \Lambda  r^2}\Bigg\}^{\frac{1}{2}} \ ,
\\\label{wt}
\Omega_\theta &=& \Omega_K\ ,
\\\label{wf}
\Omega_{\phi} &=& \Omega_K\ .
\end{eqnarray}

\begin{figure*} \centering \includegraphics[width=0.4498\linewidth]{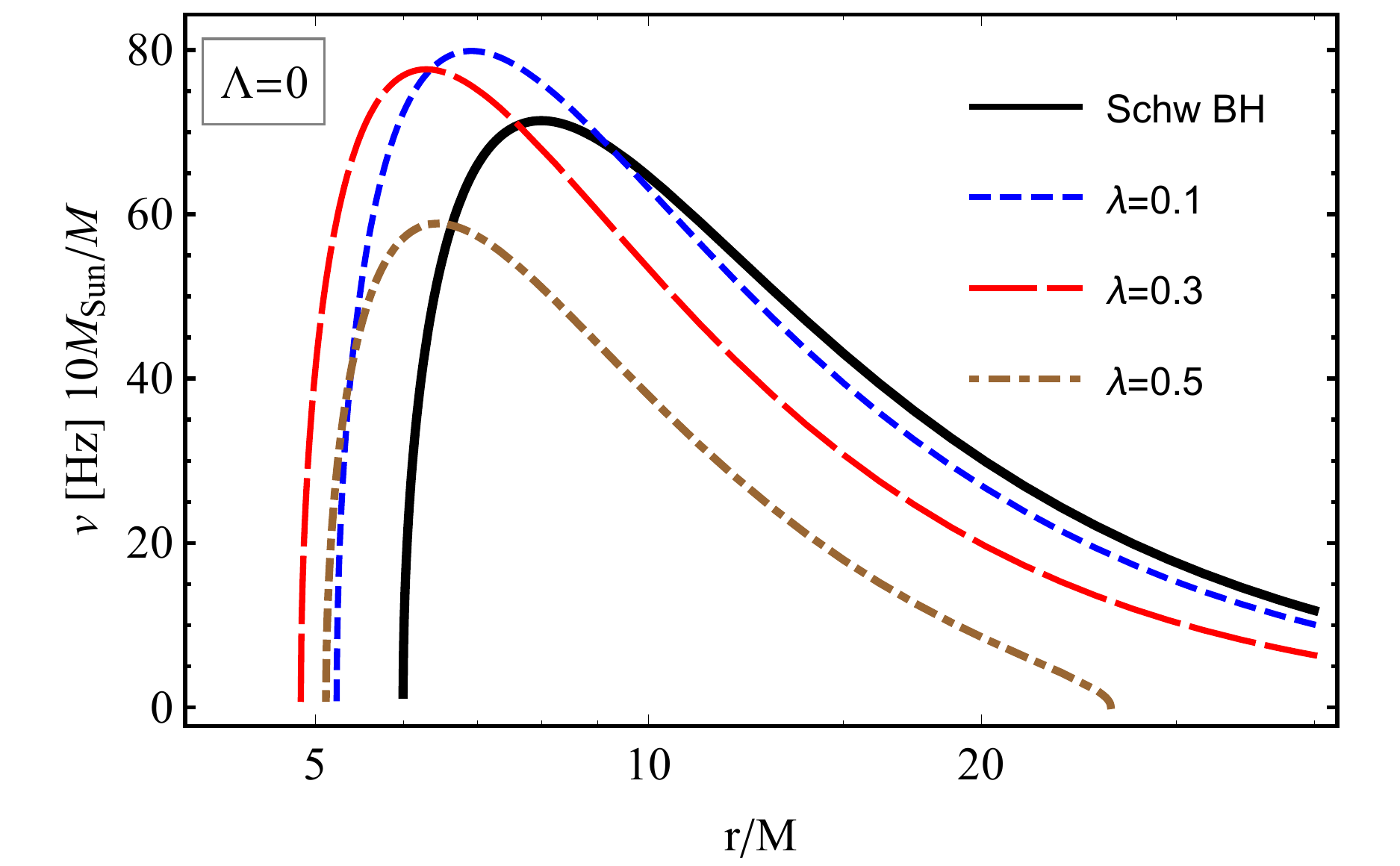}
\includegraphics[width=0.4498\linewidth]{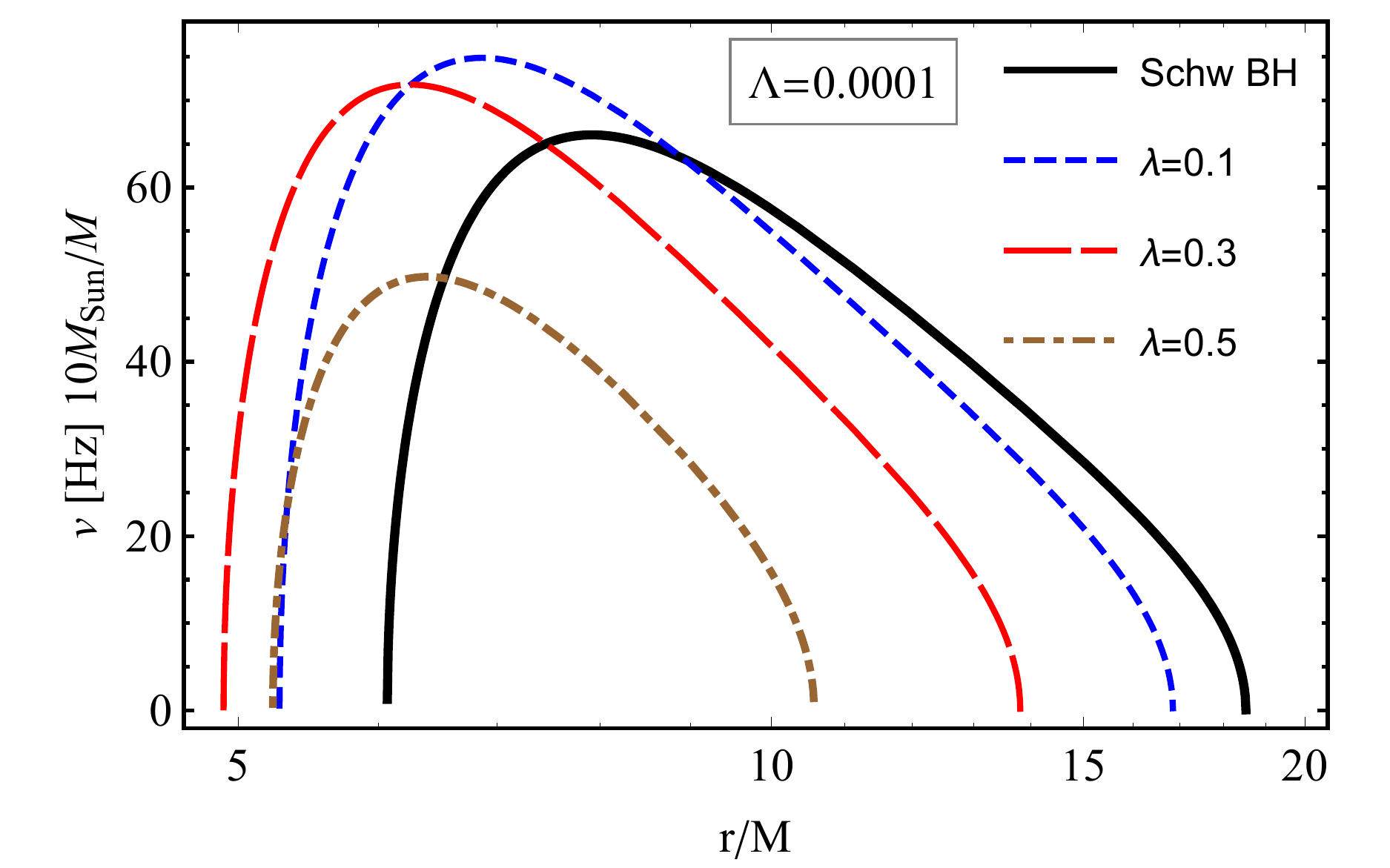}
\caption{Radial profiles of frequencies of radial oscillations of test particles around pure Schwarzschild (left panel) and Swcharzschild-de-Sitter (right panel) black hole immersed in perfect fluid dark matter field for different values of $\lambda$. In this plot we use $\Lambda \to \Lambda/M^2$ and $\lambda \to \lambda/M$.  \label{radialfreq}} \end{figure*}

Fig.~\ref{radialfreq} reflects the radial dependence of radial frequencies of test particles around Schwarzschild black hole on the left panel and around Schwarzschild-dS black hole on the right panel for different values of dark matter $\lambda$. From Fig.~\ref{radialfreq}, it can be seen that the radial frequency increases and location of oscillation shifts towards left to small $r$ due to the effect of dark matter. The inclusion of cosmological constant does however cause to the decrease of the frequency due to its cosmological repulsion. 

\subsection{Schwarzschild-de-Sitter BH in PFDM  vs Kerr BH: the same frequencies of twin peak QPOs}

In this subsection, we consider the possible frequencies of twin peak QPOs around Schwarzschild-dS black hole surrounded by perfect fluid dark matter field and compare the obtained results with the one in Schwarzschild and Kerr black hole cases for twin peak QPOs \cite{Stuchlik2016AA}:
\begin{itemize}
\item  Relativistic procession (RP) model~\cite{Stella1999ApJ}.  In the standard RP model the upper and lower frequencies are identified through the radial and orbital frequencies as $\nu_U=\nu_\phi$ and $\nu_L=\nu_{\phi}-\nu_r$, respectively. In modified RP1 and RP2 models the frequencies can be identified as $\nu_U=\nu_{\theta}$, $\nu_L=\nu_{\phi}-\nu_r$ and $\nu_U=\nu_\phi$, $\nu_L=\nu_{\theta}-\nu_r$, respectively.

\item The epicyclic resonance (ER2, ER3, ER4) models~\cite{Abramowicz2001AA}. In the ER models the accretion disk is assumed to be thick enough and QPOs appears due to the resonance oscillations of uniformly radiating particles along geodesic orbits. The upper and lower frequencies for ER2, ER3, and ER4 models are defined  through frequencies of orbital and epicyclic oscillations as $\nu_U=2\nu_\theta-\nu_r$, $\nu_L=\nu_r$,  $\nu_U=\nu_\theta+\nu_r$, $\nu_L=\nu_\theta$, and  $\nu_U=\nu_{\theta}+\nu_r$,  $\nu_L=\nu_\theta-\nu_r$, respectively.   
\item The warped disk (WD) model~\cite{Kato2004PASJ,Kato2008PASJ}. In WD model the QPOs frequencies appear due to the oscillations of warped thin disk. The upper and lower frequencies are defined as $\nu_U=2\nu_\phi-\nu_r$,  $\nu_L=2(\nu_\phi-\nu_r)$ .
\end{itemize}
\begin{figure*}\centering \includegraphics[width=0.32490\linewidth]{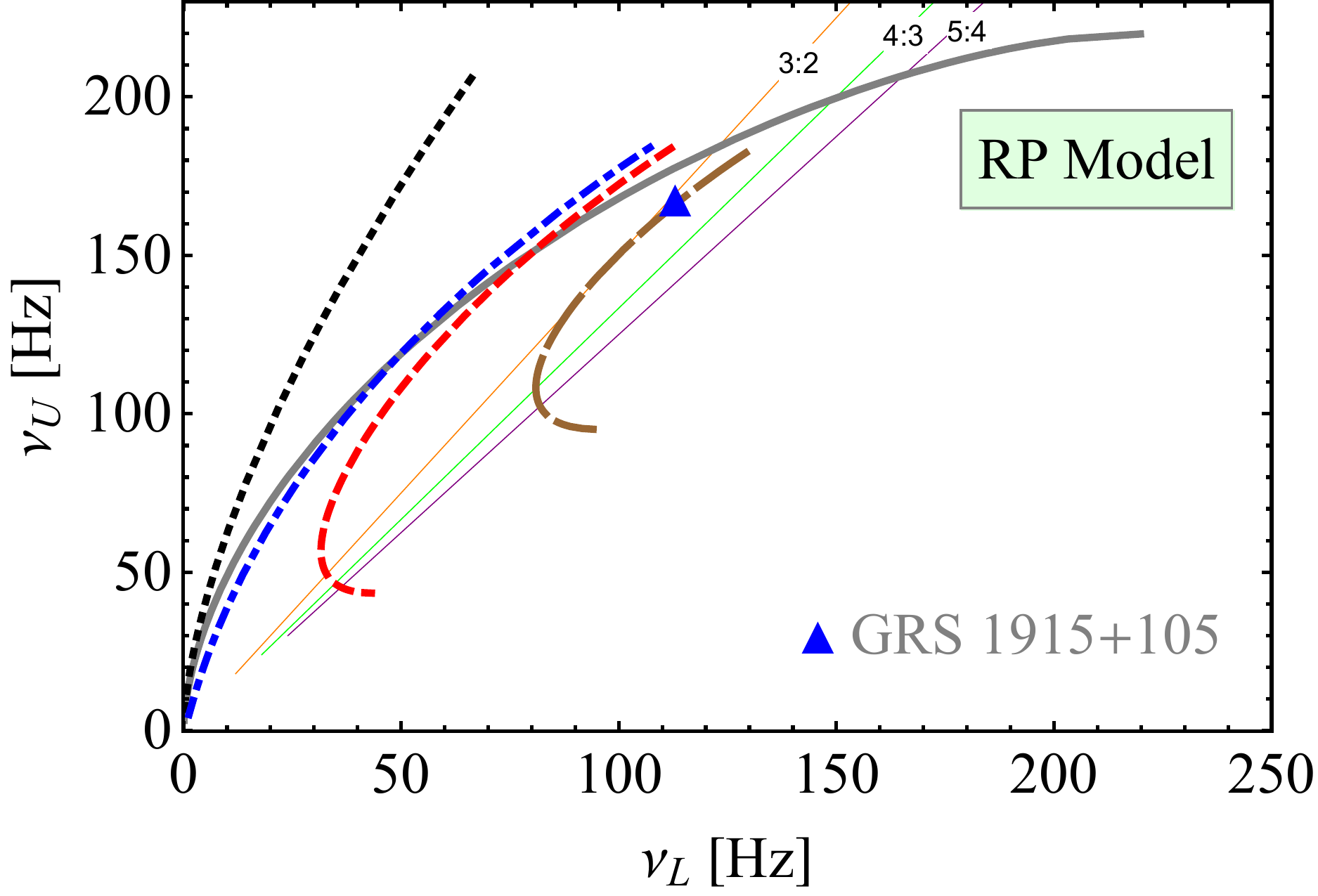}
\includegraphics[width=0.32470\linewidth]{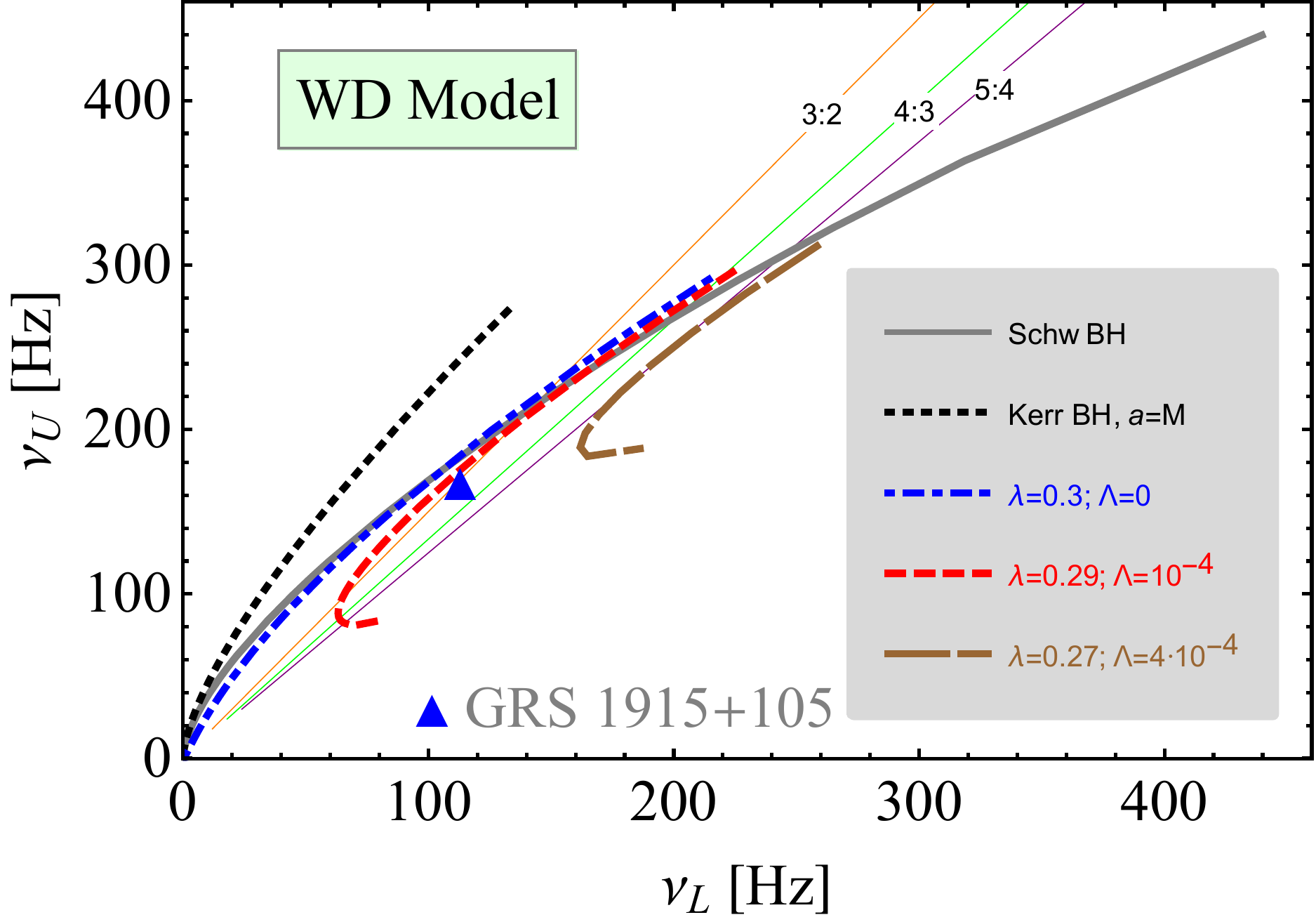} 
\includegraphics[width=0.32490\linewidth]{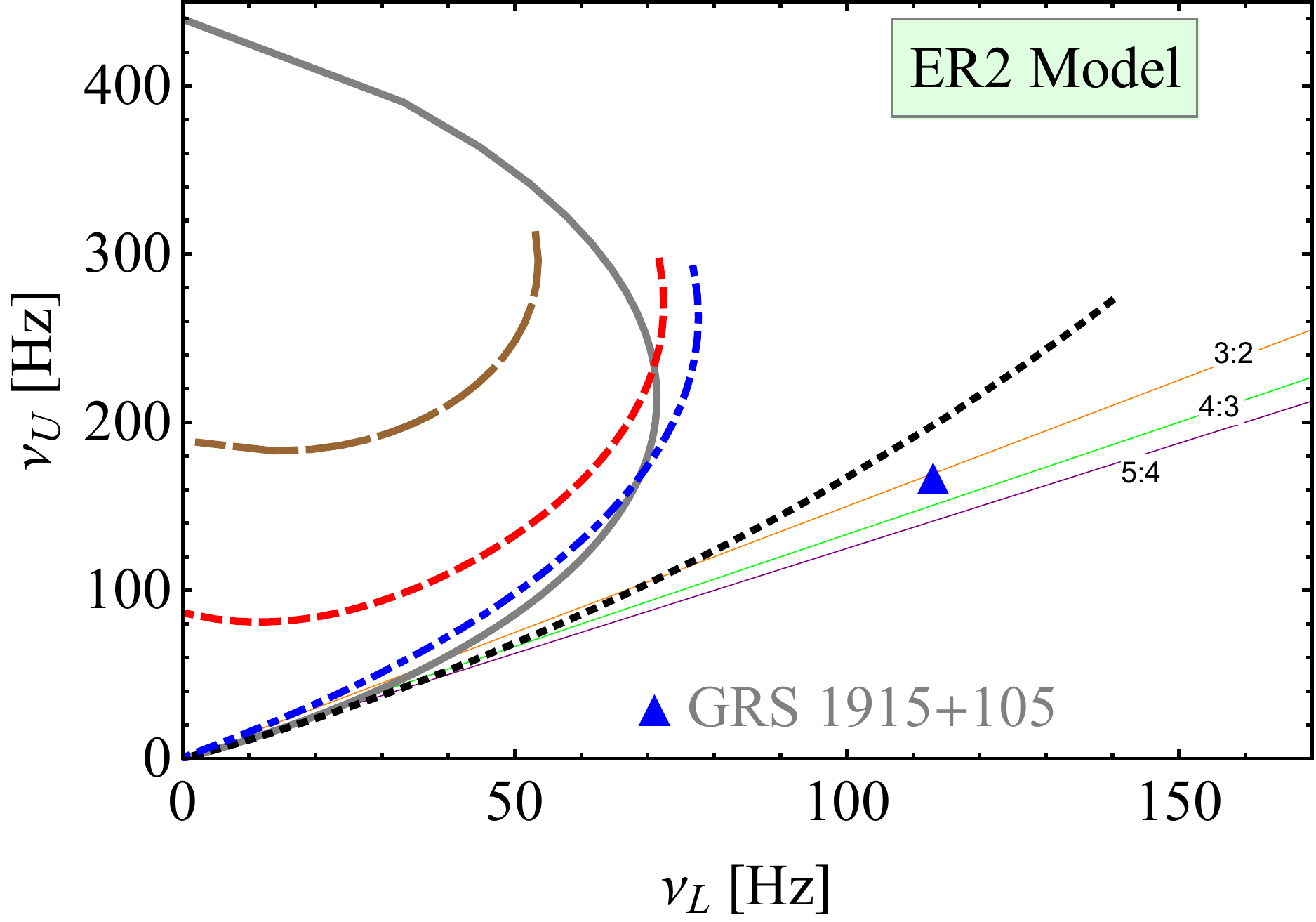} 
\includegraphics[width=0.32490\linewidth]{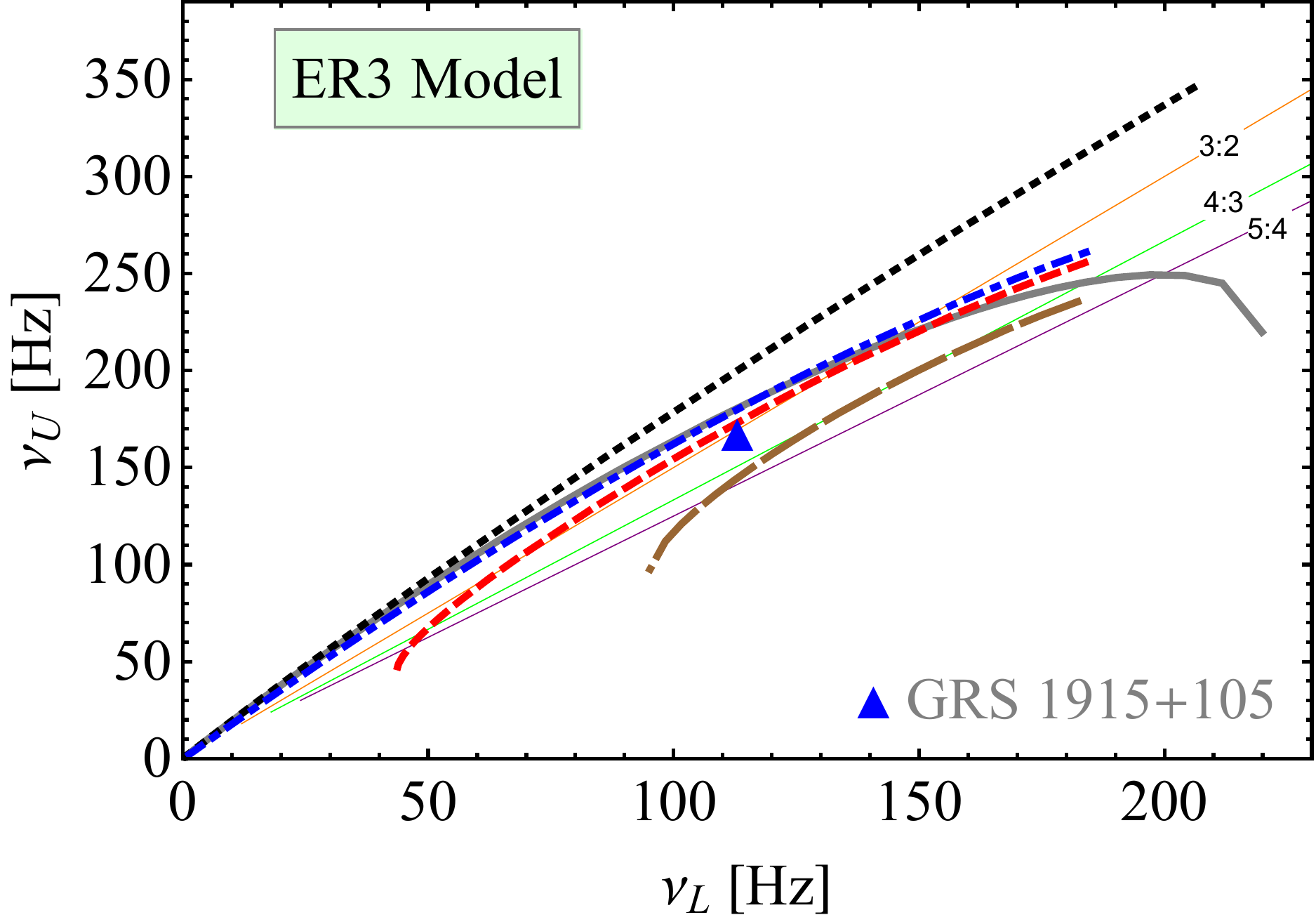} 
\includegraphics[width=0.32490\linewidth]{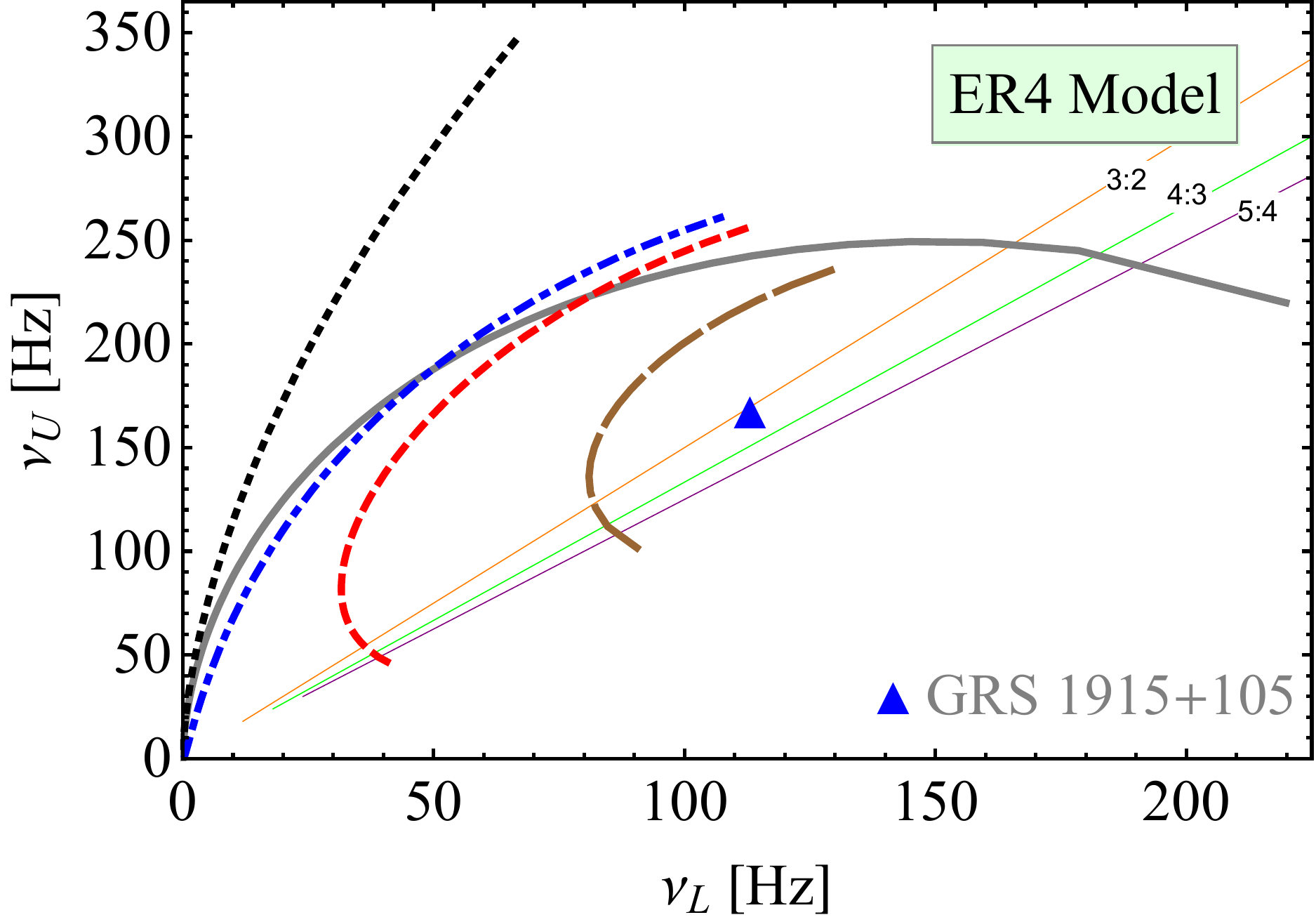}
\caption{Relations between frequencies of upper and lower peaks of twin peak QPOs in the RP, WD and ER2-4 models around Schwarzschild BH, extremely rotating Kerr BH and Schwarzschild-de-Sitter BH in PFDM for different values of cosmological constant and fixed value of PFDM parameter. In this plot we use $\Lambda \to \Lambda/M^2$ and $\lambda \to \lambda/M$. \label{QPO}} \end{figure*}

Let us then turn to the analysis of the possible frequencies of twin peak QPOs. In Fig.~\ref{QPO}, we show relations between upper and lower frequencies of twin peak QPOs around Schwarzschild blak hole, the extremal Kerr black hole, and Schwarzschild-dS black hole immersed in perfect fluid dark matter for various RP, WD and ER2-4 models. Here we focus on the set of possible values of upper and lower frequencies, and then we show the way allowing to test gravity theories on the basis of data by twin peak QPO presented in the $\nu_U-\nu_L$ diagram. According to the model here considered, if the location of a twin peak QPO lies between the black dotted and blue dot-dashed lines in $\nu_{\rm U}-\nu_{\rm L}$ space, as shown in Fig.~\ref{QPO}, a black hole can be referred to as a rotating Kerr black hole. Similarly, if the QPO lies between gray and blue dashed lines and between red dashed and brown large-dashed lines as well, there exists the effect of cosmological constant. Moreover, corresponding orange, green and purple lines representing frequencies of twin peak QPOs with the ratio of upper and lower frequencies $\nu_U:\nu_L=$ 3:2; 4:3; 5:4 suggest that the possible range of twin peak QPOs can be generated around a black hole. As can be seen from the behaviour of the twin peak QPOs the presence of cosmological constant gives rise to the decrease in the value of lower QPO frequencies up to values below 50 Hz (100 Hz) in contrast to the Schwarzschild black hole where it is greater than 120 Hz in RP, WD and ER3-4 models, respectively.

Moreover, based on the comparison of blue dashed and solid gray lines, one may infer that the possible values of upper and lower frequencies of twin peak QPOs decrease as a consequence of the presence of the dark matter. In the case of RP, WD and ER4 models, the degeneracy values of lower frequency matching with two different upper frequencies occur as that of cosmological constant. Here we restrict our attention to more realistic case that we show, as an example,  the position of a twin peak QPO referred to as GRS 1915+105 object \cite{Torok2011AA} with upper and lower frequencies $\nu_U=168\pm 5$ and $\nu_U=113 \pm 3$ Hzs, respectively and the total mass $M \sim 10 M_\odot$\ in $\nu_{\rm U}-\nu_{\rm L}$ space to find corresponding values of $\lambda$ and $\Lambda$ with the location of the QPO. Consider the numerical example: the QPO object GRS 1915+109 respectively takes place at $r/M=6.3365$ with corresponding values of $\lambda=0.24877$ and $\Lambda=4\cdot10^{-4}$ for RP model, $r/M=7.97378, 7.89477$ with $\lambda=0.363198, 0.371066$ and $\Lambda=10^{-4}$ for WD and ER3 models, and $r/M=6.87144$ with $\lambda=0.321248$ and $\Lambda=5\cdot 10^{-4}$ for ER4 model. However for ER2 model, there exists no QPO object with the frequency around Schwarzschild-dS black hole surrounded by perfect fluid dark matter.

\section{Conclusion \label{Sec:conclusion}}
 
 We studied circular orbits for test particles around Schwarzschild-de Sitter (dS) black hole surrounded by perfect fluid dark matter. We showed that the the presence of dark matter parameter reduces the ISCO radius. Similarly the unstable photon orbits decrease, depending only dark matter $\lambda$. It was interestingly shown that as a consequence of the presence of cosmological constant $\Lambda$ stable circular orbits can only exist between the ISCO from inside and OSCO from outside in the vicinity of Schwarzschild-dS black hole surrounded by perfect fluid dark matter. Howbeit, the presence of dark matter $\lambda$ does play an important role between gravitational attraction and cosmological repulsion, thus shrinking the region where circular orbits can exist. Also we found that for specific lower and upper values of dark matter parameter for the fixed $\Lambda=0.0008$ there exist double matching values for the inner and outermost stable circular orbits for which particles can have only one circular orbit as that of coincidence of the ISCO and OSCO. It is due to the fact that the dark matter contribution increases the strength of the gravitational attraction, thereby letting it to override cosmological repulsion due to $\Lambda$.   
 
We found corresponding values of $\lambda$ and $\Lambda$ for which they can cancel each other out only at a specific radius, i.e. test particle around the geometry here considered can have the same orbit as the one around the Schwarzschild geometry. From observational point of view, it would be possible to distinguish black hole geometry considered here  from other a spherically symmetric black hole. 

Further, we studied fundamental frequencies of test particles in the background geometry of Schwarzschild-dS black hole immersed in perfect fluid dark matter, i.e. Keplerian (orbital) frequency, frequencies of radial and latitudinal oscillations with their applications to upper and lower frequencies of twin peak QPO objects for various RP, WD and ER2-4 models here considered. We have shown that under the combined effects of cosmological constant and dark matter the Keplerian frequency slightly decreases.  Finally, we have explored high-frequency twin peak QPO object (GRS 1915+105) observed with upper and lower frequencies and found specific values for $\lambda$ and $\Lambda$ to best fit $\nu_U:\nu_L=$ 3:2; 4:3; 5:4 resonance of high frequency twin peak QPOs.  These theoretical studies would help to test theories of gravity by observational data from the resonance of high frequency twin peak QPOs.

\section*{Acknowledgments}
JR and SS acknowledge the support of the Uzbekistan
Ministry for Innovative Development. J.R. thanks to the ERASMUS+ project 608715-EPP-1-2019-1-UZ-EPPKA2-JP (SPACECOM).

\appendix

\bibliographystyle{apsrev4-1}  
\bibliography{references}

\end{document}